\documentclass[10pt]{article}

\usepackage{graphicx}
\usepackage{epstopdf}
\usepackage{xcolor}

\makeatletter

\def\@saveprimitive#1#2{\let#2#1}
\makeatother

\usepackage{amsmath}
\usepackage{amssymb}
\usepackage{dsfont}
\usepackage{amsthm}

\usepackage{cite}
\usepackage{caption}
\usepackage{subcaption}
\usepackage{tikz}
\usepackage{comment}

\newtheorem*{remark}{Remark}

\begin{document}
\title{
 Why is the estimation of metaorder impact \\ with public 
 market data so challenging?}
\author{%
 Manuel~Naviglio$^{1,2}$, Giacomo Bormetti$^{3}$, Francesco Campigli$^{1,4}$, German Rodikov$^{5}$,  Fabrizio~Lillo$^{1,5}$\\
 \vspace{3mm}\\
$^{1}$Scuola Normale Superiore, Pisa, Italy\\
$^{2}$ INFN Sezione di Pisa, Largo Pontecorvo 3, I-56127 Pisa, Italy\\
$^{3}$ Universit\`a di Pavia, Italy\\
$^{4}$ Universit\`a di Firenze, Italy\\
$^{5}$Dipartimento di Matematica, Universit\`a di Bologna, Bologna, Italy.\\
 }

\maketitle

\begin{abstract}

Estimating market impact and transaction costs of large trades (metaorders) is a very important topic in finance. However, using models of price and trade based on public market data provide average price trajectories which are qualitatively different from what is observed during real metaorder executions: the price increases linearly, rather than in a concave way, during the execution and the amount of reversion after its end is very limited. We claim that this is a generic phenomenon due to the fact that even sophisticated statistical models are unable to correctly describe the origin of the autocorrelation of the order flow. We propose a modified Transient Impact Model which provides more realistic trajectories by assuming that only a fraction of the metaorder trading triggers market order flow. Interestingly, in our model there is a critical condition on the kernels of the price and order flow equations in which market impact becomes permanent.  
\end{abstract}

\section{Introduction}
Transaction cost analysis is a fundamental aspect of financial trading and market impact is the main source of costs for medium and large sized investors~\cite{Bucci2019AreTI}. Thus, estimating the potential impact and cost of a trading decision is important to assess its profitability. This is particularly true and challenging for metaorders, i.e. sequences of orders and trades executed gradually over a long time period and following a single investment decision. In fact, while there is a vast literature on estimating and modeling impact of individual trades (or orders) from public data, it is less clear if and how such models can be used to estimate the expected price trajectory of a metaorder and the associated impact cost. To this end, the industrial practice is to estimate market impact and the associated cost of a metaorder  by using data on actual metaorder execution (for academic researches using this approach, see, for example, \cite{almgren2005direct,toth2011anomalous,bacry2015market,zarinelli2015beyond}).
However this approach presents some pitfalls. First of all, data on metaorder executions are typically proprietary and not publicly available. They also reflect the trading style of the specific trading firm owning the data and the conditions that lead the firm to trade. Thus they are potentially biased and very specific and might not be of general use. Second, market impact measurements are very noisy, thus large datasets are needed to accurately estimate costs. These datasets might not be available to the firm or they might require some aggregation and renormalization techniques (for example, multiple assets, long time periods, etc.) which could deteriorate the quality of the estimation.  \\

The obvious alternative is the use of publicly available market data. However in order to use them, one needs to postulate a market impact model, and then use it to estimate, analytically or via numerical simulations, the price trajectory and impact costs of a metaorder. In this case the critical aspect is the choice of the appropriate model, since its misspecification could lead to wrong  and misleading results. The advent of machine learning and artificial intelligence methods allows to use models which can be calibrated on market data and should be flexible enough to avoid the possible model misspecification. An interesting and challenging question is thus if the two approaches, the one based on actual metaorder executions and the one based on market data and impact models, provide compatible results. Although this is a very general (and maybe too ambitious) question, in this paper we provide a preliminary answer. Specifically, we consider linear and non-linear market impact models of the {\it joint} dynamics of price and trades, which we calibrate on public market data. We then use a generalized Impulse Response analysis to evaluate the average price dynamics during the execution of a metaorder predicted by the calibrated models. \\

We find that different market impact models give price profiles which are {\it not} compatible with the main stylized facts observed during the execution of real metaorders, namely the concavity of the price trajectory during the execution and a sharp convex reversion after the end of the metaorder. On the contrary, the trajectories display an almost linear increase and very small price reversion. In some cases, the predicted price trajectory even continues to grow (for a buy metaorder) for a while after the end of the metaorder.  Interestingly similar results are recently reported \cite{cont2023limit,elomari2024microstructure} using different models spanning from Vector Autoregressive (VAR) models to Generative Adversarial Networks. Specifically, in the first part of this paper we confirm these observations by considering the linear Transient Impact Model (TIM) of \cite{Bouchaud2003FluctuationsAR} and a nonlinear model based on a combination of a Convolutional Neural Network and a Long Short Term Memory calibrated on market data. For the first model, we observe that, after the calibration on data, the parameters in the infinite lags limit tend to a critical condition. This was already observed in~\cite{elomari2024microstructure} and, as discussed in the paper, could have a role in the reproduction of stylized facts of metaorders. For the Hasbrouck Structural VAR model~\cite{hasbrouck1991measuring}, we are able to provide close form expressions of the expected price trajectory. Even if with some differences, the qualitative findings of the neural network based model are similar. \\

The second part of the paper proposes a possible origin for this general disagreement and introduces a simple linear model which is capable to provide more realistic price trajectories. Starting from the TIM, we conjecture that the above-mentioned disagreement is due to the fact that statistically calibrated impact models overestimate the positive triggering effect that a child order has on the order flow of the market. As detailed in Section \ref{sec:explanation}, the explanation of this overestimation is based on the model of correlated order flow proposed by Lillo-Mike-Farmer (LMF) in 2005 \cite{lillo2005theory} and recently carefully empirically verified by Sato and Kanazawa \cite{sato2023inferring}. The LMF model proposes that the empirically observed long memory of order flow is mostly due to the superposition of trading execution of metaorders of very heterogeneous size. If this is the correct explanation, strongly autocorrelated models of order flow will be fitted on empirical data, but the addition of a child order of a new metaorder does not trigger in reality new orders with the same direction (as the statistical model would suggest) because these orders will be essentially executed according to the trading program decided beforehand.\\

In order to quantitatively elaborate this hypothesis, we introduce a modified version of the TIM where a parameter  $\alpha$  tunes the effect that a child order has on price with respect to its effect on market order flow. Using the continuous time version of the model, we are able to provide closed form expressions for the price dynamics when the kernel of the TIM and of the correlated order flow are exponential functions. For power-law kernels, which are more in line with empirical observations, we provide numerical evidences based on the discrete time version of the model and confirm the qualitative behavior of the exponential case. We find that the model has a critical condition such that after the end of the metaorder the price reverts to a constant value, i.e. despite the transient nature of the impact, there is a non zero permanent impact. In particular, as we approach this condition, as suggested by the previous calibration of the model on data, we observe that incorporating the entire metaorder contribution into the order flow causes price trajectories to become linear during the execution of the metaorder. After the metaorder ends, prices only partially revert, eventually stabilizing around a fixed value. In the exact critical case, the system exactly converges to a nonzero value corresponding to a permanent impact. Thanks to the presence of $\alpha$ we are able to reduce these effects and reproduce the empirical evidences.\\

The paper is organized as follows: Section \ref{sec:stylized} reviews the empirical literature on price dynamics during a metaorder execution. Section \ref{sec:models} introduces the problem and the estimation method based on a modified Impulse Response Analysis, providing also a close form expression for linear impact models. Section \ref{sec:empirical} presents some empirical results of the price trajectory inferred from public data with linear and nonlinear models and Section \ref{sec:explanation} provides a possible explanation of the discrepancy between empirical data and model predictions. In Section \ref{sec:modifiedTIM}, we introduce a modification of the TIM which is able to reproduce the empirical results and finally in Section \ref{sec:conclusions} we draw some conclusions.

\section{Stylized facts on the price dynamics during and after a metaorder execution}\label{sec:stylized}

We remind that a metaorder is generically defined as a sequence of orders and trades following a single investment decision. The orders composing a metaorder are often called child orders. As explained in the seminal paper by Kyle \cite{kyle1985continuous}, the choice of executing progressively (large) orders is dictated by the need of minimizing transaction costs and hiding the trading decision from the market. Both effects are consequence of market impact, the price reaction to orders and trades. Thus the price dynamics in the presence of a metaorder is different from those that would have been observed without it. This section synthesizes the main findings from the literature, focusing on the stylized facts observed in empirical studies of real metaorder executions. Later in the paper we will compare these stylized facts with the predictions obtained from statistical models calibrated on market data.\\

{\bf Price Dynamics During Metaorder Execution.}
One of the primary observations across multiple studies is that the price trajectory during the execution of a metaorder is, on average, a concave function of time \cite{moro2009market,bacry2015market,zarinelli2015beyond,Bershova}. This behavior indicates that the price impact grows at a decreasing rate as the metaorder progresses. This phenomenon might be attributed to the increasing market awareness of the metaorder's presence, leading to adjustments in the trading strategies of other market participants. There is no agreement on the functional form followed by the (average) price during metaroder execution, also because the precise shape could depend on the trading speed during the execution. A reasonable functional form, however, is a square root function of time. \\

{\bf Price Dynamics After Metaorder Execution.}
The post-execution phase typically exhibits a convex price trajectory, where the price partially reverts towards its pre-execution level \cite{moro2009market,bacry2015market,zarinelli2015beyond}. This behavior is commonly referred to as ``price reversion" and is an essential feature of market impact models. In fact, this behavior is compatible with transient impact models and not, for example, with models where impact is fixed and permanent (for example, the Almgren-Chriss model). The degree of reversion often depends on the size of the metaorder and the liquidity of the traded asset.

The amount of price reversion has been widely debated in the literature. Some papers (for example \cite{brokmann2015slow}) claim that there is a full reversion of the price, while others (for example\cite{zarinelli2015beyond}) find a non-zero long-term impact. The difficulty in reaching a conclusive answer depends essentially on three aspects: (i) after the end of the metaroder the price fluctuations are very large and one needs to average a huge number of trajectories to reach a statistically significant conclusion; (ii) the observed average decay is very slow, thus one needs to extend the measurement to many hours after the end of the metaorder, with an even larger increase of the fluctuations, and sometimes this requires to consider data from the following days; (iii) metaorders are often autocorrelated in time \cite{brokmann2015slow,bucci2018slow} and one needs to deconvolve the measured average price trajectories to obtain a result not explained by the correlation between metaorders. 

Despite being outside the scope of this paper, it is worth mentioning the empirically observed relation between the total price change between the beginning and the end of the meta-order and its volume. This is the celebrated {\it square-root impact law} \cite{toth2011anomalous,zarinelli2015beyond}, which says that the total price change is well described by the square root of the suitably rescaled metaorder volume.

Several theoretical models have been proposed to explain these stylized facts~\cite{farmer2013efficiency,Donier2014AFC}. On the contrary, in this paper we are interested  in assessing whether {\it econometric or machine learning} models, suitably calibrated on real data, can be used to obtain price trajectories that share the same properties of those observed in empirical data during the execution of real metaorders. 

\section{Price impact models and generalized Impulse Response Analysis}\label{sec:models}

A general price and trade model describes their joint dynamics, possibly by taking into account other covariates, such as limit order book quantities (spread, bid-ask volume imbalance, etc). Thus, a general model in discrete time can be written as
\begin{eqnarray}\label{eq:general_model}
p_t&=&F_1(p_{t-1},p_{t-2},..., v_t, v_{t-1}, v_{t-2},..., x_t,  x_{t-1}...), \nonumber \\
v_t&=&F_2(p_{t-1},p_{t-2},..., v_{t-1}, v_{t-2},..., x_t, x_{t-1}...), 
\end{eqnarray}
where $t\in \mathbb{N}$ is the trade time, $p_t$ is the midpoint price and $v_t$ is the signed volume\footnote{As usual, $v_t>0$ ($v_t<0$) for a buyer (seller) initiated trade. In the following, we will also indicate $v_t$ as the order flow.}
of the trade at time $t$. Notice that the definition of price at time $t$ is far from trivial and will be discussed below. Finally, $x_t$ is a vector that contains the covariates of the dynamics. 
The two functions $F_1$ and $F_2$ must be estimated from empirical data, either by using a parametric specification (e.g. a linear model, see next subsection) or approximated in a non-parametric form via, for example, neural networks or other machine learning tools. \\

{\bf Impulse Response Analysis for metaorders.} In order to estimate the price trajectory during and after a metaorder execution, we propose a modified Impulse Response Function (IRF) 
analysis. We remind that Ref. \cite{hasbrouck1991measuring} originally suggested to use the standard IRF of linear VAR model to estimate the effect on price of a{\it single trade} placed at time $t$. This is defined as
\begin{equation}\label{irf}\begin{split}
\text{IRF}(h; \delta_v)  =  ~ & \mathbb{E}[\Delta p_{t+h}|\Delta p_t, v_t+\delta_v,\Delta p_{t-1},v_{t-1},\dots]-\mathbb{E}[\Delta p_{t+h}| \Delta p_t, v_t, \Delta p_{t-1}, v_{t-1},\dots],
\end{split}\end{equation}
where the shock on volume is given by $\delta_v$ and the number of periods (horizons) $h = 0, 1, \dots, H$ are measured in ``trade time". Furthermore, to evaluate the cumulative effect of a trade in price  at each $h$, one defines the Cumulative IRF (CIRF) as $\text{CIRF}(h;\delta_v):= \sum_{h'=0}^h \text{IRF}(h';\delta_v)$. The CIRF at $h=0$ is the instantaneous impact and, under stationary models, the CIRF converges to an asymptotic value called permanent impact when $h \rightarrow \infty$. As a consequence, when $h$ is sufficiently large, the effect of $\delta_v$ on $\Delta p_t$ disappears and thus there is no longer an impact.

We propose here to use a generalized IRF analysis for assessing the impact of a metaorder. Let us consider a metaorder executed between time $t=0$ and $t=T$ and let $\delta_v$ the volume of each child trade of the metaorder. We assume that this volume is added to the market volume, thus at each time when the metaorder is traded the total volume is $v_t+\delta_v$.

When $h\le T$ the generalized IRF is defined as 
\begin{equation}\label{irf_gen}\begin{split}
\text{IRF}_\text{gen}(h; \delta_v)  =  ~ & \mathbb{E}[\Delta p_{h}|\Delta p_{h-1}, v_h+\delta_v,\dots, \Delta p_{0},v_{1} + \delta_v, \Delta p_{-1},v_{0}, \dots ] \\& -\mathbb{E}[\Delta p_{h}| \Delta p_{h-1}, v_{h}, \Delta p_{h-2}, v_{h-1},\dots].
\end{split}\end{equation}
while for $h>T$ it is 
\begin{equation}\label{eq:genIRF}
\begin{split}
\text{IRF}_\text{gen}(h; \delta_v)  =  ~ & \mathbb{E}[\Delta p_{h}|\Delta p_{h-1}, v_h,\dots, \Delta p_{T-1}, v_{T}+\delta_v, \dots \Delta p_{0},v_{1} + \delta_v, \Delta p_{-1},v_{0}, \dots ] \\& -\mathbb{E}[\Delta p_{h}| \Delta p_{h-1}, v_{h}, \Delta p_{h-2}, v_{h-1},\dots].
\end{split}\end{equation}
As we detail below, $\text{IRF}_\text{gen}(h;\delta_v)$ can be computed analytically for linear models, while it is estimated via calibrated numerical simulations for nonlinear models. In the following, to improve readability, we present the average trajectories as  $\text{CIRF}_\text{gen}(h;\delta_v)$ as a function of time $h$.\\

{\bf The definition of price increments.} The definition of the IRF requires a choice of price increments $\Delta p_t$. 
\begin{figure}[!htb]
\centering
  \includegraphics[width=12cm]{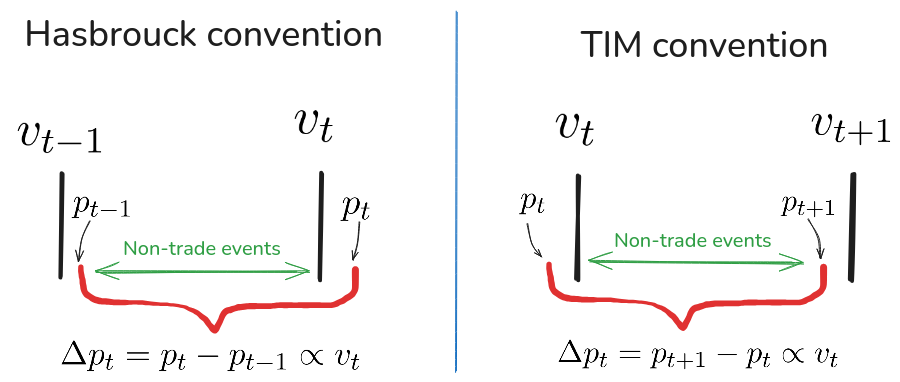}
  \caption{The figure shows the two possible conventions that can be adopted to define the price variation at time $t$.}
  \label{fig:Var_conv}
\end{figure}
In the literature, two main conventions are commonly found, as illustrated in Fig.~\ref{fig:Var_conv}. The first one, used, for example, in Hasbrouck's seminal paper \cite{hasbrouck1991summary}, defines $p_{t-1}$ as the midprice just after the trade \( v_{t-1} \). The price then evolves due to limit order depositions and cancellations until another trade occurs at time \( t \), and immediately after the midprice is \( p_t \). In this convention, the price contains information about the state of the Limit Order Book (LOB) immediately after the trade, and lacks information about the state of the LOB just before the trade. Moreover the definition of price change contains the limit orders and cancellations arriving before the trade and not in reaction to it. Thus the contemporaneous relation between trade volume and price change reads in this convention
\begin{equation}\begin{split}\label{eq: Var_H}
   \Delta p_t = p_t - p_{t-1} \propto f(v_t),
\end{split}\end{equation}
where $f$ is a generic function (linear in linear models).

In the second convention, used for example in the TIM\cite{Bouchaud2003FluctuationsAR},  $p_t$ is the midprice immediately before the trade at time \( t \) with volume $v_t$. After this trade, the LOB evolves due to limit orders and cancellations until the instant of time just before the trade at $t+1$ where the midprice is  \( p_{t+1} \). Thus, in this case, the price \( p_{t+1} \) incorporates not only the immediate reaction of the LOB to the trade at time $t$, but also its subsequent evolution due to non-trade events triggered by the trade itself. In this case the indexing convention for the immediate impact is different from the previous one and it is 
\begin{equation}\begin{split}\label{eq: Var_TIM}
    \Delta p_t = p_{t+1} - p_{t} \propto f(v_t).
\end{split}\end{equation}
In comparing Eq. (\ref{eq: Var_H}) and (\ref{eq: Var_TIM}) we observe a different definition of $\Delta p_t$ (a part from a different definition of $p_t$ itself). Although this difference is important when estimating these models on real data, we expect that when considering long time horizons, as, for example, in metaorder execution, the two conventions should lead to similar results.

\subsection{Linear models}

We consider first a linear specification of the general model of Eq.~\eqref{eq:general_model}. This model is essentially\footnote{We choose the price increment and the volume, instead of the price return and the trade sign indicator, as in the original model, because we want to estimate the dynamics of price and we are interested in the impact of traded volumes, rather than accumulated trade signs.} the SVAR model proposed by Hasbrouck in~\cite{hasbrouck1991summary} and defined by
\begin{equation}\label{hasbrouckmodel}
\begin{split}
\Delta p_t = \sum_{i=1}^{p}a_i \Delta p_{t-i} + \sum_{i=0}^{p}b_i v_{t-i} + u_{1,t}, \\
v_t = \sum_{i=1}^{p}c_i \Delta p_{t-i} + \sum_{i=1}^{p}d_i v_{t-i} + u_{2,t}.
\end{split}
\end{equation}

The terms $u_{i,t}$ are uncorrelated white noises with given distributions that captures all price changes not directly due to trades. 
Defining $y_t=(\Delta p_t, v_t)'$, the model can be written in a compact form as
\begin{equation}\begin{split}\label{svarP}
y_t =  \sum_{i=1}^{p} \tilde{A}_{i} y_{t-i}  + \tilde{u}_t,
\end{split}\end{equation}
where $\tilde{A}_i$ are suitable matrices and $\tilde u_{t}$ are correlated noises.
SVAR model parameters can be easily estimated using ordinary least squares (OLS) since there are no identifiability problems due to the well defined causal structure between the variables (trades simultaneously impact prices but not the other way around). We name this model simply as H($p$). Furthermore, as any VAR$(p)$ model, Eq.~\eqref{svarP} can be written in the companion form as 
\begin{equation}\begin{split}\label{eq: companion}
    z_t = \Gamma z_{t-1} + \epsilon_t,
\end{split}\end{equation}
where $z_t$ and $\epsilon_t$ are two $2p$ dimensional vectors and $\Gamma$ is a suitable $2p\times 2p$ matrix.

It is worth noticing that a special case of the H($p$) model is the Transient Impact Model, obtained by setting to zero the $a_i$ and $c_i$ terms, namely
\begin{equation}\label{transientimpact}
\begin{split}
\Delta p_t =  \sum_{i=0}^{p}b_i v_{t-i} + u_{1,t}, \\
v_t =  \sum_{i=1}^{p}d_i v_{t-i} + u_{2,t}.
\end{split}
\end{equation}

The standard IRF of Eq. \ref{irf} for a SVAR($p$) Hasbrouck model can be written as (see \cite{lutkepohl2005new})
\begin{equation}\begin{split}\label{irf}
    \text{IRF}(h;\delta_v) = J \Gamma ^h J' (0,\delta_v)',
\end{split}\end{equation}
 where $J := [I_2:0: \cdots:0]$ is a $2 \times 2p$ selection matrix and $I_2$ is the $2\times 2$ identity matrix. The linearity of the SVAR models implies that the IRF in Eq. \eqref{irf} depends only on the estimated parameters and not on the process history and that it is linear on the ``size" of the shock $\delta_v$.

The generalized Impulse Response Function of Eq. \ref{eq:genIRF} can be written in a compact form for a linear model. In Appendix~\ref{GIRA}, we show that the expected midprice $\tilde{p}_k$ during the metaorder can be expressed as the cumulative sum of the average price increments as
\begin{equation}\begin{split}\label{pk}
    \tilde{p}_k   = p_0 + \sum_{j=1}^{k}\Delta \tilde{p}_j  =  p_0 + \delta_v \Big[ e_1' (I_{2p} - \Gamma)^{-1}\times (k I_{2p}  - \Gamma (I_{2p}-\Gamma)^{-1}(I_{2p}-\Gamma^k))e_2\Big],
\end{split}\end{equation}
where $\Gamma$ is defined in Eq.~\eqref{eq: companion}, $I_{2p}$ is the $2p\times 2p$ identity matrix while $e_1$ and $e_2$ are $2p \times 1$ vectors defined as $e_1:=(1,0,\dots,0)'$ and $e_2:=(b_0,1,0,\dots,0)'$. Here, since the relation is intended under unconditional expectation, $\tilde z_t = \mathbb{E}[z_t]$.
Furthermore, for a metaorder of length $T$, we obtain that the expected price dynamics after the execution is given by
\begin{equation}\begin{split}\label{pTkCorp}
    \tilde p_{T+k} = \tilde p_T + \sum_{i=1}^{k} \Delta \tilde p_{T+i} = \tilde p_T + e_1' \Gamma (I_{2p}
-\Gamma)^{-1} (I_{2p}-\Gamma^k) \tilde z_T.
\end{split}\end{equation}

Given the price dynamics before (Eq.~\eqref{pk}) and after (Eq.~\eqref{pTkCorp})  the metaorder execution, in Appendix~\ref{GIRA} we derive the conditions required to obtain concave growth and convex relaxation of the price, as observed in real data.

\section{Empirical results}\label{sec:empirical} 

Here we present the results on the price dynamics during and after the execution of a metaorder according to the above introduced models with parameters calibrated on real data. 

We use data obtained from LOBSTER\footnote{lobsterdata.com}, containing observations from NASDAQ-listed stocks during the 22 trading days of June 2021. Specifically, we examine Microsoft, as representative of large tick stocks, and Amazon, as a small tick stock and the price changes are computed in the Hasbrouck convention. As customary, we exclude the first and last 30 minutes to reduce the effect of intraday pattern. Furthermore, we merge orders occurring at the same timestamp, with nanosecond precision, and having the same sign.

\subsection{Results for the linear model}
\label{sec:results}

To estimate the parameters of the SVAR and reduced models of Equations~\eqref{hasbrouckmodel} and~\eqref{transientimpact}, respectively, we use Ordinary Least Squares (OLS). This procedure mitigates possible estimation issues due to a misspecification of the distribution of the residuals. The choice of the number of lags \( p \) in linear models is critical to accurately capture the dynamics of price and volume, as well as the properties of the price dynamics around a metaorder. Originally, Hasbrouck \cite{hasbrouck1991measuring} suggested setting \( p \) equal to five. However, subsequent studies have indicated that such setting might be insufficient to fully describe the properties of order flows and signed volumes, which exhibit characteristics consistent with long-memory processes \cite{Bouchaud2003FluctuationsAR,lillo2004long,bouchaud2009markets,toth2015equity,bouchaud2018trades}. Moreover, an interesting recent contribution \cite{elomari2024microstructure} finds that the largest eigenvalue of a VAR model describing price and market variables tends to one when the number of lags of the VAR is increased.
To balance the effort of capturing long-memory effects and managing model complexity, we evaluated the model at two large $p$ settings: \( p = 2 \times 10^{3} \) and \( p = 4 \times 10^{3} \). These settings were selected to probe the depth of memory effects and the feasibility of parameter estimation, thereby enhancing the model's ability to reflect the underlying dynamics accurately.

\begin{figure}[!htb]
\centering
\includegraphics[width=8cm]{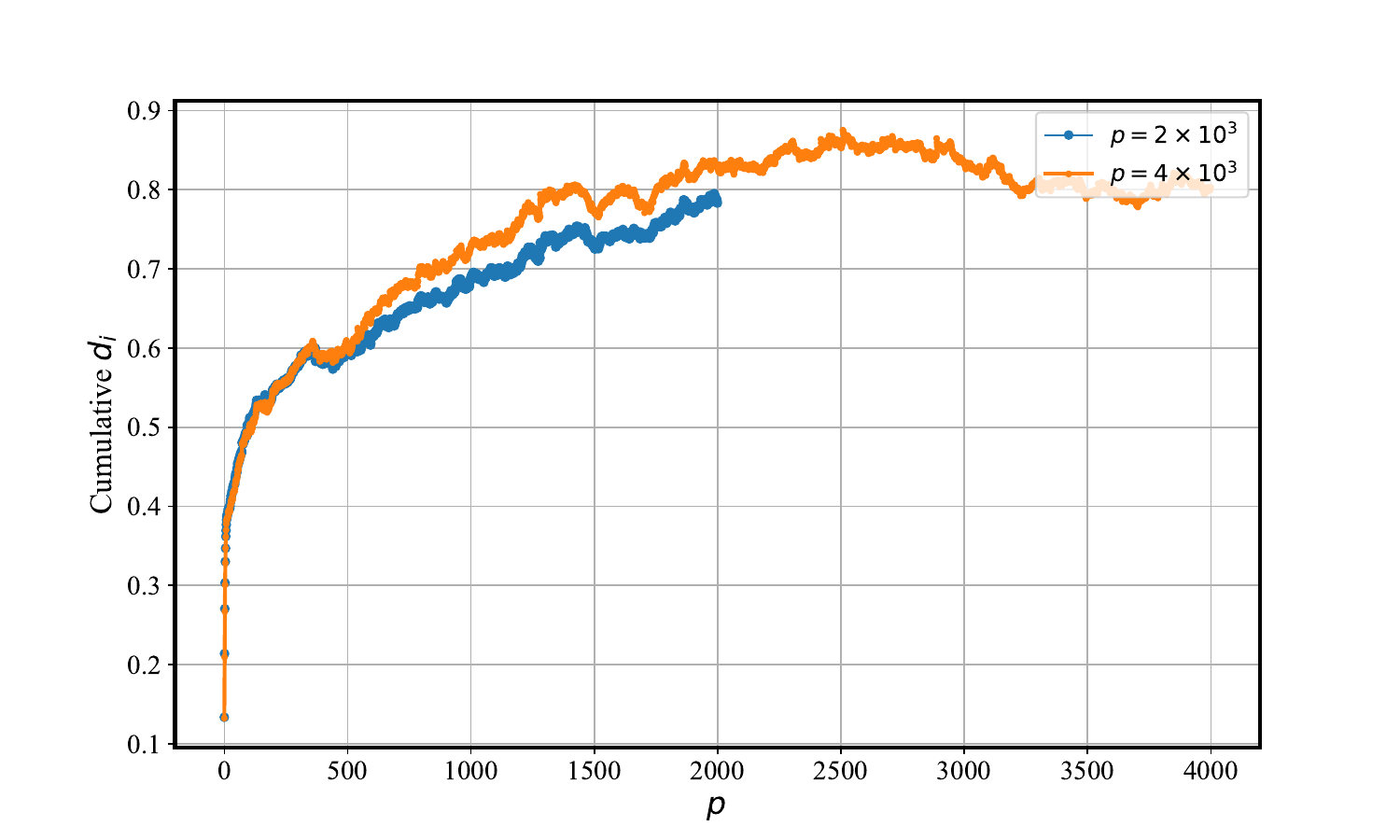}
\includegraphics[width=8cm]{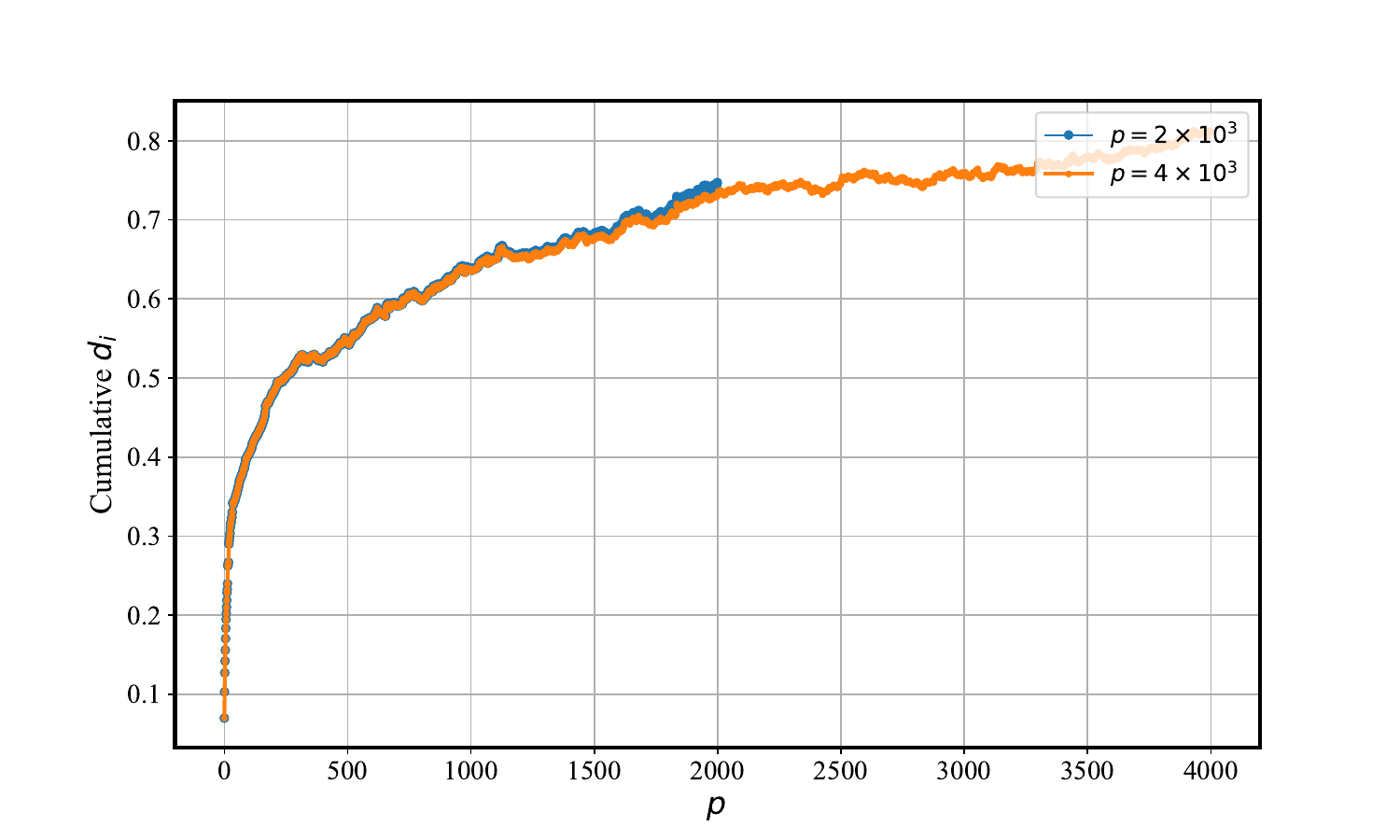}
\caption{Cumulative $d_i$ parameter plots with \( p = 2 \times 10^{3}\) and \( p = 4 \times 10^{3}\) for Amazon (left panel) and Microsoft (right panel).}
\label{fig:diParsCritical}
\end{figure}

Upon reviewing the cumulative coefficients from our parameter estimations, it became evident that within the Hasbrouch model of Eq. (\ref{hasbrouckmodel}), the coefficients \( a_i \) and \( c_i \) do not manifest significant cumulative effects, especially when contrasted with \( b_i \) and \( d_i \). 
Consequently, we redirected our focus towards the more influential coefficients, leading us to further investigate the TIM\footnote{Anyhow, qualitative similar results for the price trajectory are obtained for the full Hasbrouck model.} of Eq.~(\ref{transientimpact}).
We remind that in an AR(p) model as the second equation in (\ref{transientimpact}) the sum of the \( d_i \) coefficients indicates the level of endogeneity of the process and when this sum approaches the value $1$ the process becomes non stationary. Figure 
\ref{fig:diParsCritical} shows this cumulative sum for the two considered values of the order $p$. In both cases it is evident that, increasing the order of autoregression, the cumulative sum of the \( d_i \) coefficients becomes closer and closer to 1 and this suggests that the volume process is approaching non-stationarity, consistently with the recent findings in \cite{elomari2024microstructure}.

When we add the volume of the metaorder execution, the equations of the  model become 
\begin{equation}
\label{eq:transient_impact_metaorder_with_metaorder_lambda}
\begin{split}
\Delta p_t &= \sum_{i=1}^{p} b_i v_{t-i} + b_0 v_{t}, \\
v_t &= \kappa \left( \sum_{i=1}^{p} d_i v_{t-i} \right) + \delta_{v}.
\end{split}
\end{equation}
where $\delta_v$ is the trade size of each child order of the metaorder and \( \kappa = \{0,1\}\) is a dummy variable. The case $\kappa=0$ corresponds to the case where market trading volumes are described by a white noise and the child order of size $v_t=\delta_v$ does not affect the market trading volume. On the contrary, when $\kappa=1$ the autoregressive dynamics of volumes is taken into account and the child order of size $\delta_v$ influences market volume which is then transferred to prices. We arbitrarily set the volume $\delta_{v}$ of each child order of the metaorder equal to the median trade size of the considered asset, which is 7 shares for Amazon and 100 shares for Microsoft. Note however that, because of the linearity of the model, the impact is proportional to the child order size.  The metaorder lasts from time $t=0$ to \( t = T=1000 \) trades.

\begin{figure}[!htb]
\centering
\includegraphics[width=8cm]{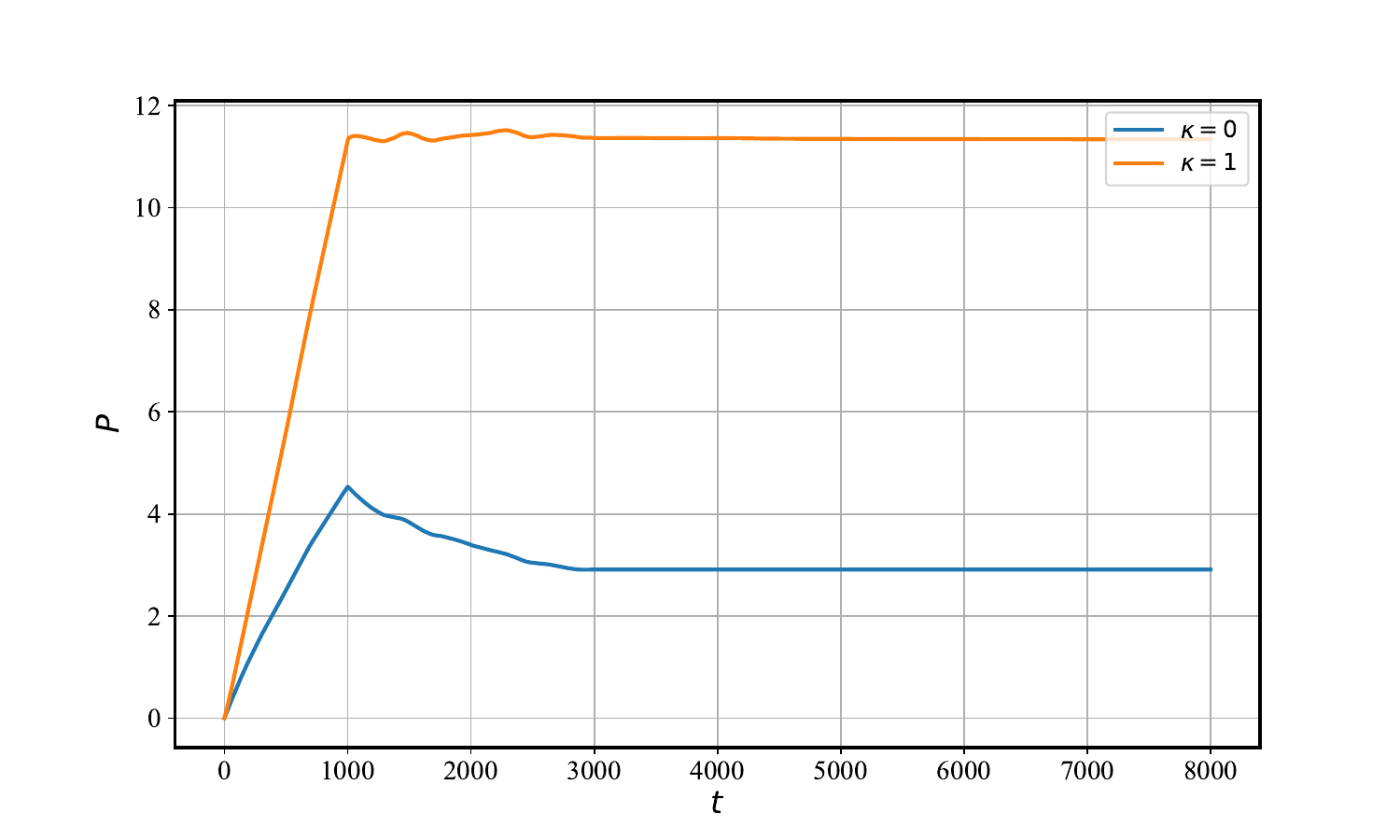}
\includegraphics[width=8cm]{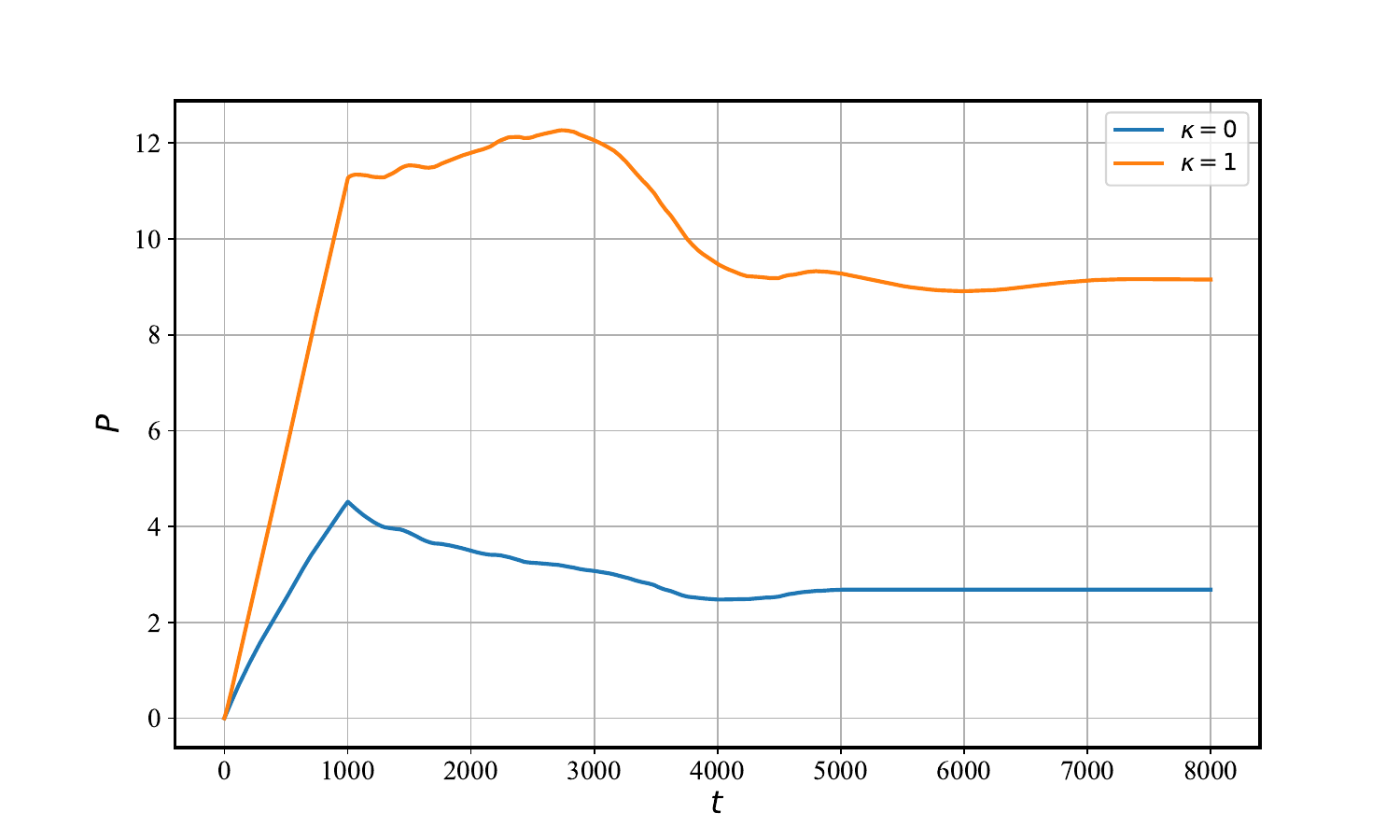}
\caption{Simulated price trajectory with respectively $p= 2 \times 10^{3} $ (left panel) and $p = 4 \times 10^{3}$ (right panel), for Amazon featuring metaorder execution for $T = 1000$.}
\label{fig:simulated_prices_comparison_amzn}
\end{figure}

\begin{figure}[!htb]
\centering
\includegraphics[width=8cm]{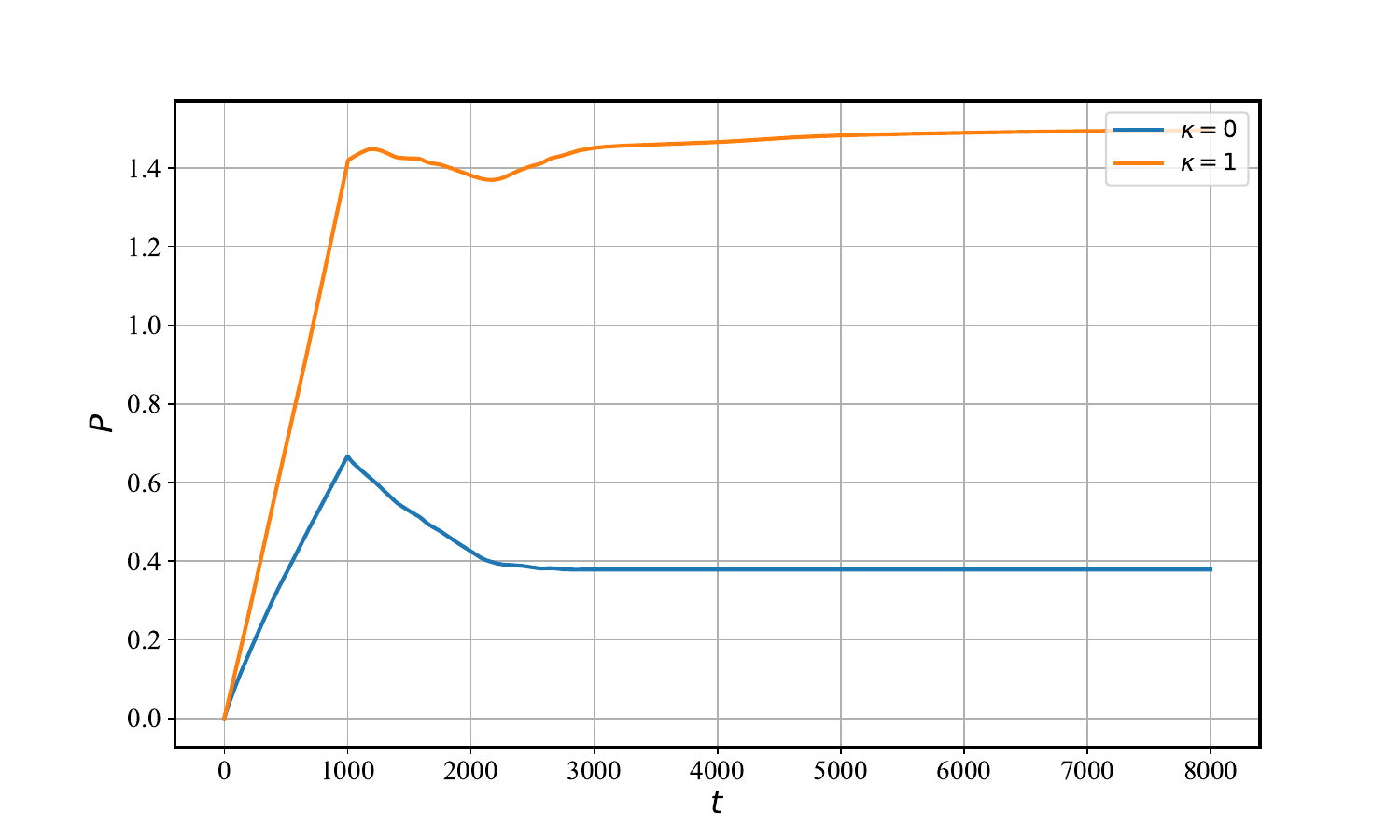}
\includegraphics[width=8cm]{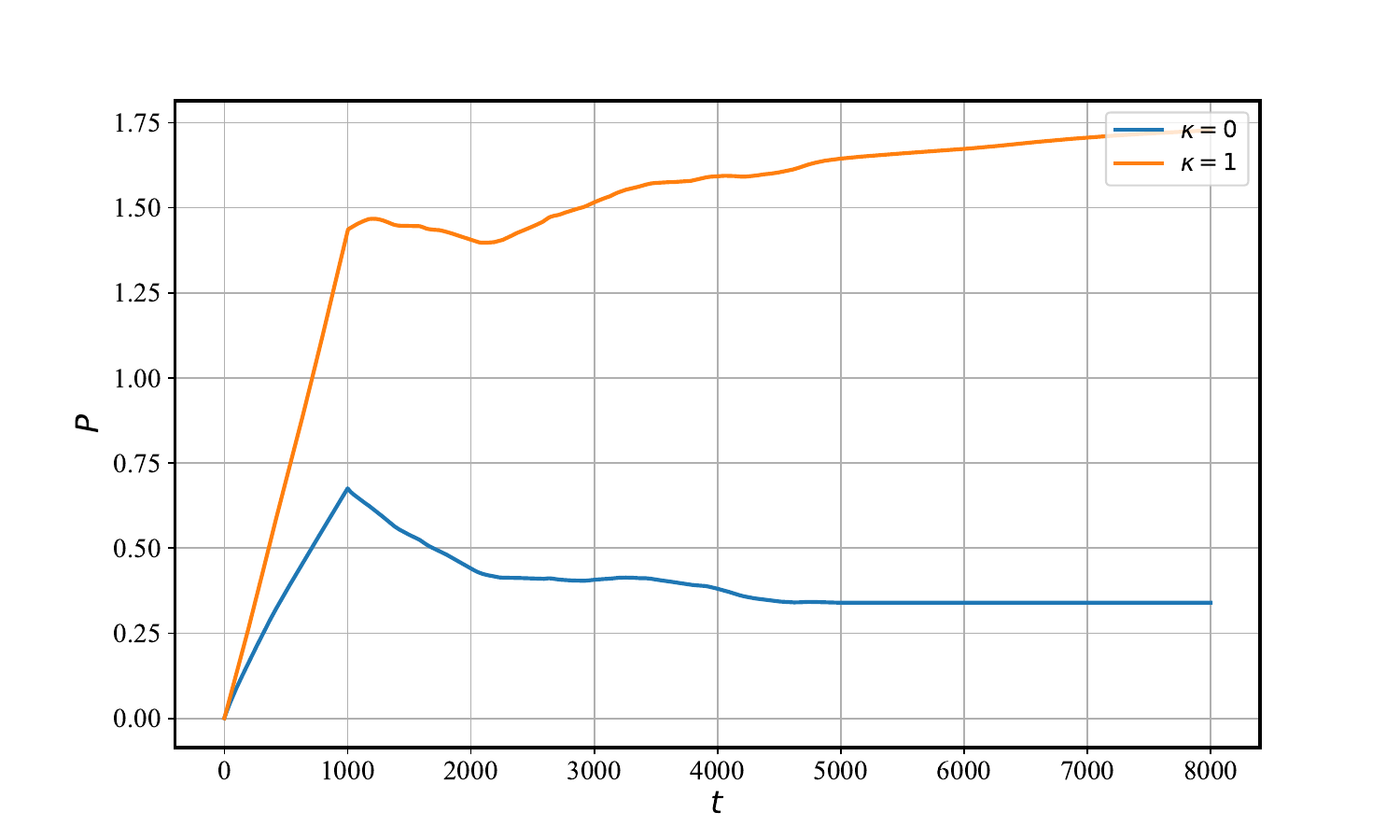}
\caption{Simulated price trajectory with respectively $p= 2 \times 10^{3} $ (left panel) and $p = 4 \times 10^{3}$ (right panel), for Microsoft featuring metaorder execution for $T = 1000$.}
\label{fig:simulated_prices_comparison_msft}
\end{figure}

Figures~\ref{fig:simulated_prices_comparison_amzn} and~\ref{fig:simulated_prices_comparison_msft} show the results for the estimated price dynamics during and after the metaorder execution when $p=2\times 10^3$ (left panels) and $p=4\times 10^3$ (right panels) form AMZN and MSFT, respectively. When considering the model without the autoregressive dynamics for the volume ($\kappa=0$), we observe a mildly concave dynamics of the price during the execution of the metaorder and a partial reversion of the price after its end. Moreover, the amount of price reversion after the metaorder is larger for larger $p$ and the price is going to be completely reverted if $p\to \infty$ (i.e. if $p\gg T$). This in line with what known for the TIM model, see for example the discussion in \cite{bouchaud2009markets}.

When \( \kappa = 1 \), i.e. when the metaorder volume has an effect on market volume, the price impact dynamics is importantly modified. 
As seen from the figures, the price dynamics during the execution of the metaorder is essentially linear (or even mildly convex) and there is no appreciable reversion after its end. Moreover, and importantly for transaction cost analysis, the peak impact $p_T-p_0$ is roughly twice larger than in the case $\kappa=0$. The reason is that the metaorder triggers other market trades more likely with the same sign as the metaorder, pushing the price even higher (for a buy metaorder) than in the case with $\kappa=0$. 
Similar patterns are reported in~\cite{cont2023limit,elomari2024microstructure}, suggesting similar behavior under comparable conditions.

Interestingly, when increasing $p$ things get worse, in the sense that price patterns looks less similar to the ones obtained with real metaorder. Looking at the right panel of Fig. \ref{fig:simulated_prices_comparison_amzn} for Amazon, a sort of pronounced inertia effect is observed: after the end of a buy metaorder, the price continues to increase for a significant amount of time before starting to decline. The explanation is that even when the metaorder is over, the order flow triggered by the metaorder is significant, also due to the large value of $p$ and sustains the increase of price. This effect is less marked in Microsoft (right panel of Fig. \ref{fig:simulated_prices_comparison_msft}), potentially reflecting differences in tick size and the slope of cumulative \( d_i \) coefficients. These observations suggest the hypothesis that tick size and microstructural differences may influence how quickly trade impacts dissipate, but more extensive analyses are needed.

These findings highlight that when jointly modeling  the price and trade volume dynamics, i.e. $\kappa =1 $ in the above notation, the coefficients capture a much longer memory effect. This causes prices to increase almost linearly with time, to relax much more slowly than expected, leading, in some cases, to what we called the inertia effect. Thus, one of the possible reasons for the recent empirical observations reported in Ref.s~\cite{cont2023limit,elomari2024microstructure} using different and more complex models, could be related to the activation of long-memory volume dynamics triggered by metaorders shocks. In particular, the contribution of $\delta_v$ on volume dynamics in our simulations seems to trigger much stronger effects on prices than expected leading to clear differences from what observed on real metaorder data. In contrast, for $\kappa=0$, for which $\delta_v$ only influences price dynamics, we recover the concave and convex behaviour of prices similar to what empirically found.
These observations suggest that only a fraction of the metaorder shock should contribute on volume dynamics while the rest couldbe directly incorporated in price dynamics. This guess is investigated in the Section \ref{sec:modifiedTIM}, where we introduce a modified version of the continuum TIM model in which the metaorder contribution can be split in the two equations. 

\subsection{Results for a non-linear LSTM model}\label{sec:lstm}

One might question whether the unexpected results we observed are due to the linear nature of the TIM, since it might lack the complexity needed to capture the intricacies of the market. For this reason, we perform an analogue analysis using an highly non-linear model based on Neural Networks. In particular, we consider a combination of Convolutional Neural Networks (CNN) and Long Short Time Memory (LSTM). The first ones are deep learning architectures commonly used for processing grid-like data, such as images, and can also be applied to sequential data like time series. In the context of time series, CNNs apply convolutional operations to the input sequence, enabling them to capture \textit{local} patterns and temporal dependencies within the data. The latters are another type of deep learning architecture suitable for sequential data processing. LSTMs are designed to capture temporal dependencies by maintaining hidden states that evolve over time as the network processes each element in the sequence. This enables LSTMs to model complex sequential relationships in time series data. Unlike CNNs, LSTMs have recurrent connections that allow them to maintain memory of past inputs, making them well-suited for tasks where historical context is important, such as time series predictions. The main difference between CNNs and LSTMs in the context of time series data lies in how they capture temporal dependencies. CNNs primarily capture local patterns and short-range dependencies within the data through convolutional operations, while LSTMs capture long-term dependencies by maintaining memory of past inputs through recurrent connections. The two architectures can be combined in such a way to take advantage of both properties.

Here we consider a model composed by a first 1D CNN layer with 32 filters and a $\tanh$ activation function, followed by two LSTM layers with 64 neurons. Then, a decision layer is built as a dense layer with a sigmoid activation function, whose output is again given to another LSTM with the same hyperparameters as before. Finally, a two dimensional dense layer gives the output of the model predicting the next step of the volume and price time series. Thus, the model takes as input length varying time series of trade signed volumes and price changes and gives as output the next step of each of the two time series. The model has been trained and tested (80\% and 20\%) on one month of Microsoft (June 2021) using the same dataset as in the linear model case, minimizing an MSE loss function. The length of the input time series is $5$ and 100. The model has good out of sample $R^2$ performances in both cases both for returns ($\sim 13 \%$) and volumes ($\sim 2\%$). After the calibration of the model, we simulate the market reaction to a metaorder execution by shocking the volumes with $\delta_v=100$, as done in the linear case, that are inserted as input of the volume in the Neural Network model. 

\begin{figure}[!htb]
\centering
\includegraphics[width=8cm]{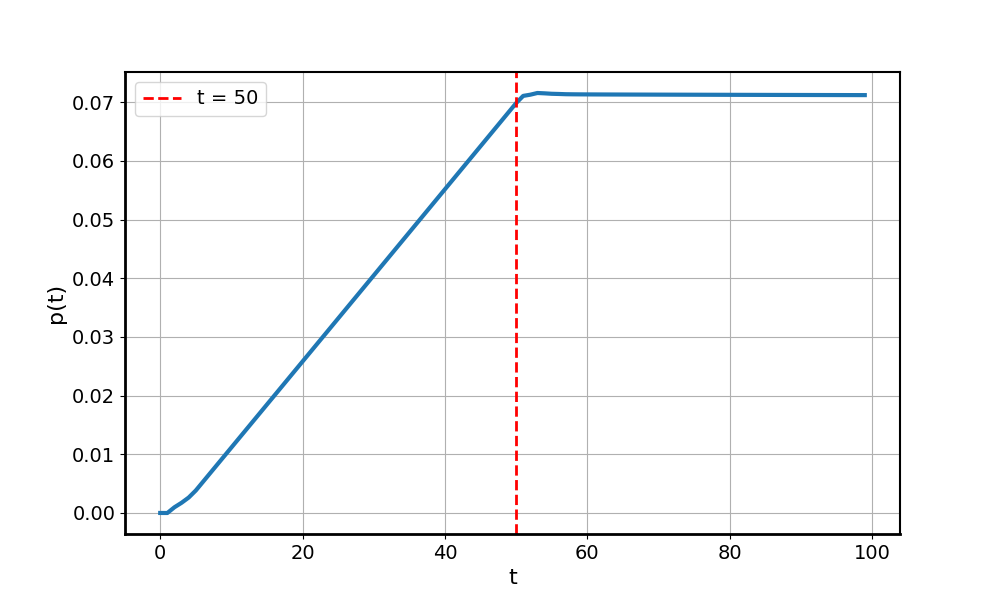}
\includegraphics[width=8cm]{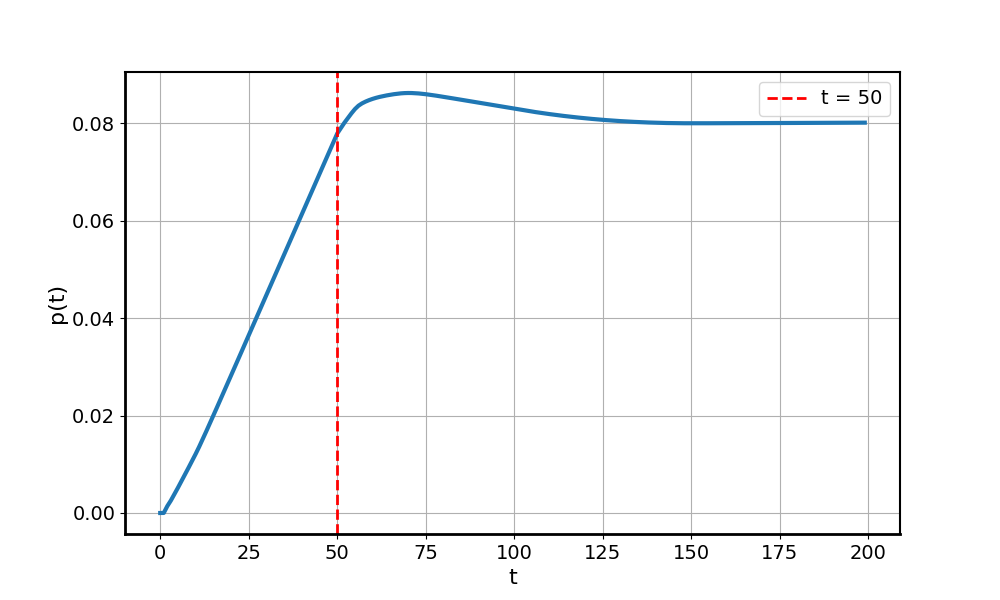}
\caption{Price dynamics during and after the execution of a metaorder of length $T=50$ using a Neural Network calibrated on Microsoft data. The left panel shows the case of 5 lags while the right one the case of 100 lags. The right panel also shows an inertia effect which makes the price increase also after the end of the metaorder execution.}
\label{fig:Metaorder_NN}
\end{figure}

Figure~\ref{fig:Metaorder_NN} shows the result of the simulation. Also in this case the behaviour of the metaorder is not reproduced by the model. In fact the price increases almost linearly during the execution and its decay after the metaorder execution is not visible.

A similar shape was already observed in~\cite{cont2023limit}, where, still using a non linear model, the expected response of the market to the metaorder execution was not reproduced. We emphasize that further studies on neural network models capable of more accurately reproducing the expected behavior are required, and this task is deferred to future work. Here, we show that behavior similar to that obtained with linear models can also be achieved using nonlinear models. These observations suggest a deeper underlying reason why metaorder execution is not well described by either linear or nonlinear models. To address this, in Section \ref{sec:modifiedTIM}, we investigate this issue by introducing a modified version of the TIM that incorporates the metaorder volumes. 

\section{Why are statistical models unable to reproduce price trajectories during metaorder executions?}\label{sec:explanation}

The puzzling results shown in the previous Sections, which are consistent with those obtained in the literature, indicate that a purely econometric/statistical model, correctly capturing the lagged correlations, is unable to reproduce the price trajectories during metaorder execution. A possible explanation could be given by considering the model proposed by Lillo, Mike, and Farmer in 2005 \cite{lillo2005theory} and recently verified empirically in Ref. \cite{sato2023inferring}. According to this model, the long range autocorrelation of order flow is due to the simultaneous execution of a certain number of {\it independent} metaorders, whose size (in number of transactions) is distributed asymptotically as a power law. In other words, each metaorder contributes independently to the autocorrelation with a time scale related to its execution time, and the aggregation of many time scales provides the observed long memory. 

In this setting, the addition of an extra metaorder would not alter the order flow of the rest of the market because the added metaorder does not modify the execution of the existing ones\footnote{To be precise, there is a small modification due to the fact that each metaorder trades now on average every $M+1$ trades rather than every $M$ trades, where $M$ is the number of metaorders. However, this modification is negligible for large $M$.}. Thus, despite the fact one econometrically measures a strong autocorrelation of order flow, the addition of a new one does not modify the order flow as it would be predicted in an Impulse-Response analysis. 

A partial and indirect empirical justification of this assumption is provided in \cite{Tóth01072012,doi:10.1142/S2382626618500028}. The first paper uses labeled trade data from the LSE allowing to identify the market member responsible of each market order. It is observed that the sign autocorrelation conditional to the fact that both trades are initiated by the same market member is way larger than the sign autocorrelation conditional to the fact that the two trades are initiated by two different market members. The second paper uses proprietary data on metaorder executed by Capital Fund Management. Also in this case, the autocorrelation of trade sign when one of the two trades is from CFM and one from the market is much smaller than the unconditional autocorrelation. This indicates that trades from a metaorder might trigger much less additional order flow than what is expected by the unconditional autocorrelation of trade sign. 

The scenario suggested by the Lillo-Mike-Farmer model is probably extreme. It is possible, in fact, that each trade only partially triggers new order flow. In the next Section we propose a simple model where each trade of a metaorder triggers in part new order flow from the market which will impact the price, and in part impacts direclty the price. In a limit we obtain the pure TIM, where the metaorder acts as an external source for the price, while in the other, similar to the coupled price and order flow model of Eq.~\eqref{eq:transient_impact_metaorder_with_metaorder_lambda} or of the Hasbrouck model, the metaorder volume fully affects the market order flow. We will show that in the intermediate cases some interesting dynamics for the price can be observed.

\section{A modified Transient Impact Model}\label{sec:modifiedTIM}

We first present the model in continuous time, because this allows us to derive a closed-form expression for the price dynamics. Later, we provide the discrete-time version, which will be useful to extract some relevant results when analytical expressions are not available. In the following we neglect for simplicity the noise contribution in the price and in the volume equations, in other words we are considering the deterministic skeleton of the dynamics. The model's equations for $t\ge 0$ are 
\begin{equation}\begin{split}\label{eq: VolterraEqs}
    & p(t) = \int_0^t G(t-\tau)\big[v(\tau) + (1-\alpha)V\theta(T-\tau)\big] d\tau, \\&
    v(t) = \alpha V \theta(T-t) + \lambda \int_0^t \mathcal{D}(t-\tau) v(\tau) d\tau.
\end{split}\end{equation}
where $p(t)$ is the midpoint price at time $t$ and $v(t)dt$ is the market signed volume in $[t,t+dt]$. We assume as initial conditions that $p(0)=v(0)=0$. As before, a metaorder is executed in $[0,T]$ and the trading rate is $V$ (thus the total volume of the metaorder is $VT$). The metaorder contribution to the volume is non-zero for \(t < T\) and zero for \(t \geq T\) and \(\theta(T-t)\) is the Heaviside step function. Additionally, we introduce the parameter, \(\alpha \in [0,1]\), which quantifies the contribution of the metaorder to the order flow from the rest of the market\footnote{One can choose two different parameters $\alpha_1$ and $\alpha_2$ not necessarily summing to one. We made this choice to have only one tunable parameter describing the different cases.}, thus it represents a trade-off between two limits: when \(\alpha = 0\) the metaorder contribution only affects the price dynamics without changing the market volume dynamics. When \(\alpha = 1\), instead, the metaorder contribution triggers the volume dynamics that will eventually contribute to the price dynamics.  Finally, the functions $G$ and $\mathcal{D}$ are given kernel functions: the first one is the propagator of the TIM, while the second describes the long-range autocorrelation of market volume. The latter is also modulated by the parameter $\lambda$ which has the dimension of the inverse of a time and, as will be clear in the following, takes the role of a characteristic scale of frequencies that influences both volume and price dynamics. Note that the case in which $\alpha=0$ has a correspondence with the case $\kappa=0$ discussed in the empirical analysis. Indeed, in both cases the metaorder contribution does not affect volume dynamics but only the price one. Thus, in the following we always consider $\lambda$ appreciably different from zero, since the case $\lambda\sim 0$ is trivial and can already be explored by taking $\alpha=0$. Note that one could add to each of Eqs~\eqref{eq: VolterraEqs} an additional noise term that takes into account the part of the dynamics not due to trades. However, this term can be incorporated into the model without any issue. Since the noise has zero mean and is uncorrelated with the trading volume and with the price, averaging the processes provides the deterministic dynamics described by the solution of Eqs~\eqref{eq: VolterraEqs}. This has been verified through numerical simulations of the discretized version of the model in the presence of noise.

The second equation in~\eqref{eq: VolterraEqs} is a Volterra integral equation of the second type, whose solution is
\begin{equation}\begin{split}\label{eq: VolterraSolution}
    v(t) = \mathcal{L}^{-1}\bigg[\frac{\mathcal{L}[f]}{1-\lambda \mathcal{L}[\mathcal{D}]}\bigg](t),    
\end{split}\end{equation}
where $\mathcal{L}$ denotes the Laplace transform and $f(t)=\alpha V \theta(T-t)$ represents the trading profile of the metaorder. By solving the Laplace transforms in Eq.~\eqref{eq: VolterraSolution}, one can determine the volume dynamics of the system and then, by simple integration, the price dynamics, $p(t)$. 

The volume equation includes the metaorder term, which temporarily perturbs the system's equilibrium. However, we expect the volume to eventually return to an equilibrium state, converging to a fixed point for $t \gg T$. To get an idea of the  qualitative behavior of the solution of the volume equation, we first consider the dynamics during the execution of the metaorder ($0<t<T$,), i.e.
$$
v_1(t) = \alpha V + \lambda \int_0^t \mathcal{D}(t-\tau) v_1(\tau) d\tau.
$$
During the execution of the metaorder $V$, volumes are non-stationary and increase over time, impacting price dynamics in a similar way throughout. Then, when considering the dynamics after the end of the metaorder, ($t>T$), the equation becomes
$$
v_2(t)=\lambda \int_0^T \mathcal{D}(t-\tau) v_1(\tau)d\tau +\lambda \int_T^t \mathcal{D}(t-\tau) v_2(\tau)d\tau \equiv g(t) +\lambda \int_T^t \mathcal{D}(t-\tau)v_2(t) d\tau
$$
For large \(t\), the function \(g(t)\) approaches zero, leading \(v_2(t)\) to also decay to zero if the following condition is satisfied:  
\begin{equation}\begin{split}\label{eq: StatCond}
    \int_0^\infty \mathcal{D}(t) \, dt < 1.
\end{split}\end{equation}  
When this condition holds, the integral operator contracts the time dependence, ensuring convergence to a fixed point. Conversely, if the integral exceeds 1, the operator amplifies the time dependence, causing the volume to diverge in the long-time limit. A particularly interesting case arises when the integral equals one, i.e.
\begin{equation}\begin{split}\label{eq: Critical}
     \int_0^\infty \mathcal{D}(t) \, dt = 1.
\end{split}\end{equation}
This corresponds to the \textit{critical case}, where the volume converges to a constant, non-zero asymptotic value, indicating that the initial volume perturbation caused by the metaorder persists indefinitely.\\Therefore, for the volume to converge to a finite fixed point at large times, the integral must either be strictly less than 1, leading to a decay to zero with no memory, or equal to 1 in the critical case, where the volume asymptotes to a finite value retaining a permanent memory.

It is worth noting that, by looking at the first Equation~\eqref{eq: VolterraEqs}, one might be misled into thinking that the contribution to the price of the volume of the child orders of the metaorder is weighted differently from the rest of the order flow, represented by $v(t)$, due to the presence of the parameter $(1-\alpha)$. This would appear to produce a distinct influence on the price, implying that the market should be able to distinguish the metaorder from the rest of the market.  However, it is easy to see that the equation can always be rewritten as
\begin{equation}\begin{split}\label{eq:PriceDecomposition}
    p(t) = \int_0^t G(t-\tau) \big[ \tilde{v}(\tau) + V(\tau) \big] \, d\tau,  
\end{split}\end{equation}
where $\tilde{v}(\tau)$ denotes the background order flow component, while $V(\tau)$ represents the metaorder contribution, more precisely
\begin{equation}\begin{split}\label{eq:vDecomposition}
    \tilde{v}(t) = \lambda \int_0^t \mathcal{D}(t-\tau)\, v(\tau) \, d\tau, 
    \qquad
    V(t) = V\,\theta(T-t).   
\end{split}\end{equation}
Equation~\eqref{eq:PriceDecomposition} is the standard representation of price dynamics, in which the total order flow can be split into a component associated with the metaorder and another describing the rest of the market. 
Thus, the representation given in Equation~\eqref{eq: VolterraEqs} is useful to identify which portion of the child orders of size $V$ contributes to the overall order flow and which portion directly affects the price dynamics. 
However, the same model can always be rewritten as in Equation~\eqref{eq:PriceDecomposition}, where it becomes clear that the background market component and the metaorder term produce exactly the same market impact.

In Appendix~\ref{Appendix_Discretisation}, we show that, in the absence of metaorders (\( V=0 \)), the continuous time model in Eq.~\eqref{eq: VolterraEqs} corresponds to either the discrete Hasbrouck model, Eq.~\eqref{hasbrouckmodel}, or TIM, Eq.~\eqref{transientimpact}, depending on whether we approximate the integral using the right or left endpoints of each subinterval in the Riemann sum, in the limit where the number of lags \( p \) in the sums extends to the entire history of the time series.

\begin{remark}
The model in Eq. \ref{eq: VolterraEqs} describes the price and order flow dynamics {\it conditional} to the presence of a metaorder. One interesting question\footnote{We thank an anonymous referee for this suggestion.} is how to generalize this model when many metaorders are present at the same time. In this case the term $ V\theta(T-t)$ is replaced by a generic stationary process $V(t)$ describing the volume coming from metaorders, while $v(t)$, as before, is the volume not coming from metaorders. According to the Lillo, Mike and Farmer model \cite{lillo2005theory}, the heterogeneity of metaorder size leads to a long-memory correlation for $V(t)$. Thus the question is under which conditions the price process of this modified model of Eq. \ref{eq: VolterraEqs} remains diffusive. 
In Appendix~\ref{Diffusivity} we prove that the discretized version of the modified stationary model satisfies the same conditions between the exponent of the propagator $G$ and the one of the autocorrelation of metaorder order flow as the standard transient impact model to yield a diffusive price process (see \cite{bouchaud2009markets}).
\end{remark}

In the next paragraph, we study this model using exponential kernels, for which an exact solution can be found. We show that even with kernels that are not long memory, such as exponential kernels, the price can preserve an impact which is as longer as more the parameters of the model approach critical conditions. In particular, near the critical condition, the trajectory of the price during the execution of the metaorder becomes linear when $\alpha=1$. Furthermore, under the same conditions, the model shows unexpected memory effects that are not in agreement with the empirically observed convex drop in prices after the end of the execution. 

\subsection{Continuous time version: exponential kernels}

In this Section, we analyze the case where the kernel \(\mathcal{D}(t)\) has the exponential form \(\mathcal{D}(t) = e^{-\beta t}\). Under this assumption, the stationary conditions~\eqref{eq: StatCond} and~\eqref{eq: Critical} impose the following constraint on the parameters:
\begin{equation}\begin{split}\label{eq: HawkesCondition}
    \beta \geq \lambda.
\end{split}\end{equation}
The strict inequality corresponds to the stationary condition~\eqref{eq: StatCond}, where volumes decay exponentially to zero. In contrast, the equality corresponds to the critical condition~\eqref{eq: Critical}, where volumes—and consequently prices—retain a permanent impact.

 In the following, we first derive the volume dynamics before and after the metaorder execution, and then the corresponding price dynamics still using an exponential kernel in the form $G(t)=e^{-\rho t}$.

\subsubsection{Volume dynamics}
In Appendix~\ref{VolterraResolution}, we explicitly derive the solution of the Volterra equation for the volume, which is for all $t$
\begin{equation}\begin{split}\label{eq: VolDynamicsMain}
    v(t) = \begin{cases} \alpha V \bigg\{ \bigg( \frac{\beta}{ \beta - \lambda} + \big(1-\frac{\beta}{ \beta - \lambda}\big) e^{-(\beta - \lambda) t}\bigg) -  \theta(t-T) \bigg( \frac{\beta}{\beta-\lambda} + ( 1-\frac{\beta}{\beta - \lambda} ) e^{-(\beta-\lambda)(t-T)} \bigg)\bigg\}, & \text{if $\beta \neq \lambda$}, \\ \alpha V \big\{  1 + \beta t  - \theta(t-T) [ 1 + \beta (t-T) ] \big\}, & \text{if $\beta = \lambda$}.
\end{cases}
\end{split}\end{equation}
It can be noticed that the $\lambda$ parameter assumes the role of an inverse time scale which is put in relation with  $\beta$, representing the inverse time scale of the exponential kernel ${\mathcal D}$. 

Although the solution holds for any $\beta$ and $\lambda$, as discussed above, condition~\eqref{eq: HawkesCondition} must be satisfied to ensure that volumes remain stationary.

\begin{figure}[!htb]
\centering
\includegraphics[width=8cm]{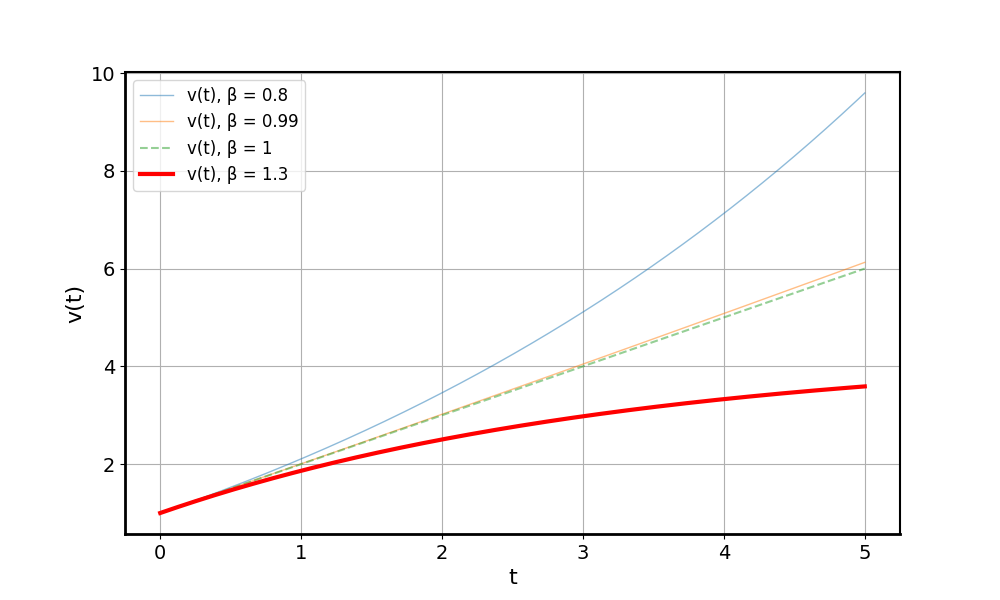}
\includegraphics[width=8cm]{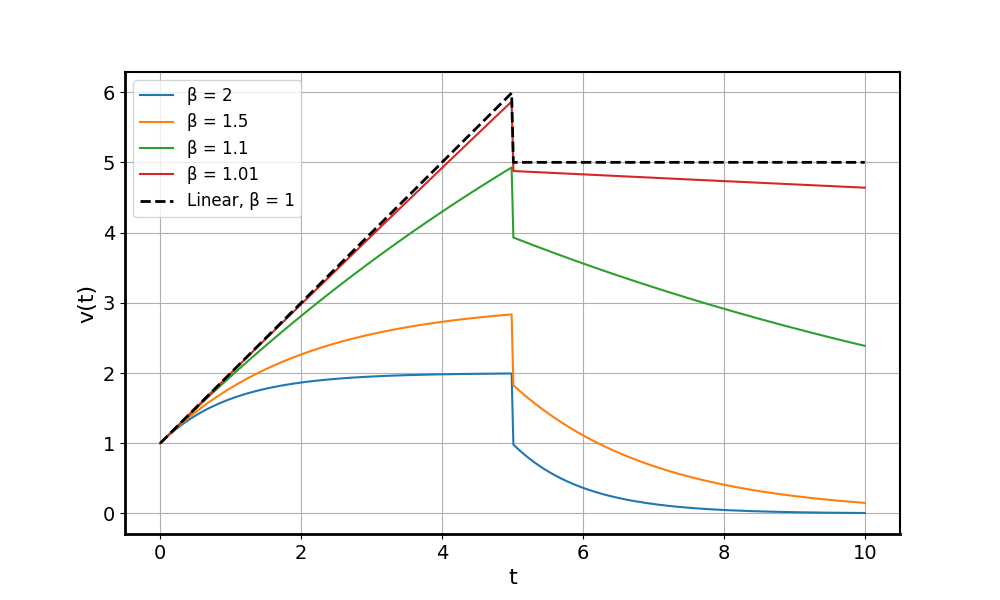}
\caption{Left. Volume dynamics for the exponential kernel case for $\alpha=1, V=1$. If we fix $\lambda$, in this case $\lambda=1$, the volumes fall into two different regimes depending on the relative magnitudes of $\beta$ and $\lambda$. Right. Dynamics of volumes during and after the metaorder. The volumes will always eventually converge to zero while for $t\sim T$ there is a discontinuity due to the sudden stop of the metaorder.}
\label{fig:vt_beta}
\end{figure}

{\bf Volume dynamics during metaorder execution.} 
The left panel of Figure \ref{fig:vt_beta} presents the volume dynamics during the metaorder, i.e. for $t<T$. When $\beta > \lambda$, the volume is described by a concave function of time, which converges to a fixed value, $v^*$,  for sufficiently long metaorders, i.e., when $T \gg 1/\beta$. In the same figure, we also show the critical case, i.e. $\beta = \lambda$, and non stationary case, i.e., $\beta < \lambda$. We observe a transition to a divergent, convex behavior in correspondence of the critical condition, where the volume exhibits a linear dependence on execution time (from Equation~\eqref{eq: VolDynamicsMain}, for $\beta = \lambda$, the volume is given by \( v(t) = \alpha V \big\{ 1 + \beta t \big\} \) for $t<T$). Therefore, when the model is close to the critical case, the volume increases quasi-linearly, and the concave convergence to $v^*$ appears only for long executions.

{\bf Volume dynamics after metaorder execution.} The volume dynamics after the end of the execution, i.e., for $t > T$, are given by
\begin{equation}\begin{split}\label{eq: VolDynamicsMainRelax}
    v(t) = \begin{cases} 
    \alpha V  \left(1-\frac{\beta}{ \beta - \lambda}\right) \left( e^{-(\beta - \lambda) t} - e^{-(\beta-\lambda)(t-T)} \right), & \text{for $\beta > \lambda$}, \\ 
    \alpha V \beta T, & \text{for $\beta = \lambda$},
\end{cases}
\end{split}\end{equation}
 The behavior of volumes, before and after the metaorder, is shown in the right panel of Fig.~\ref{fig:vt_beta}. The volume exhibits a discontinuity at $t = T$ due to the end of the metaorder. When $\beta > \lambda$, the volume exponentially decays to zero. In contrast, when $\beta = \lambda$, after the metaorder, the volume does not decrease but instead immediately converges to the constant value $\alpha V \beta T$. In this case, the volume retains a \textit{permanent memory} of the metaorder and continues to contribute consistently to the price dynamics. Additionally, the figure illustrates the case where $\beta = 1.01$ and $\lambda = 1$. Here, we still have $\beta > \lambda$. However, since we are close to the critical case, we observe a quasi-linear growth of volumes and a very slow relaxation over long timescales.

We conclude that volumes, even with an exponential kernel, can exhibit very long memory, and potentially permanent memory in the critical case $\beta=\lambda$. This is due to the presence of the time scale $\lambda^{-1}$. 
Thus, being in proximity of the critical case, i.e., $\beta \sim \lambda$, would lead to linear volume growth during the metaorder and very slow relaxation afterward. In the next paragraph, we discuss how this phenomenon could, within our model, trigger long-term memory effects even in prices which again present a linear growth during the execution and a long scale relaxation afterwards, as empirically observed after the calibration of the model on data.

\subsubsection{Price dynamics}
After computing volume dynamics, we can substitute the result in the expression of prices in Eq.~\eqref{eq: VolterraEqs} and extract the related dynamics. The computation is performed in details in Appendix~\ref{PricesDyn} and the result reads
\begin{equation}\begin{split}\label{eq: PriceDynMain1}
    p(t) &= \alpha V \bigg\{  \frac{\beta}{\beta - \lambda} \frac{1-e^{-\rho t}}{\rho} + \bigg( 1-\frac{\beta}{\beta-\lambda} \bigg) \frac{e^{(-\beta + \lambda)t  } - e^{-\rho t}}{\rho - \beta + \lambda} \\& - \theta(t-T)   \bigg( \frac{\beta}{\beta-\lambda} \frac{1 - e^{- \rho(t-T)}}{\rho}  +  ( 1-\frac{\beta}{\beta - \lambda} ) \frac{  e^{-(\beta-\lambda)(t-T)  } - e^{-\rho(t-T)} }{\rho - (\beta-\lambda) } \bigg) \bigg\} \\& + (1-\alpha) V  \bigg\{ \frac{1-e^{-\rho t}}{\rho} - \theta(t-T) \frac{1-e^{-\rho(t-T)}}{\rho}  \bigg\},
\end{split}\end{equation}
for $\beta \neq \lambda$, while it is
\begin{equation}\begin{split}\label{eq: PriceDynMain2}
p(t) & = \alpha V \bigg\{ \frac{1-e^{-\rho t}}{\rho} + \beta ~ \frac{\rho t + e^{-\rho t}-1}{\rho^2} - \theta(t-T) \bigg[ \frac{1-e^{-\rho(t-T)}}{\rho}+\beta ~ \frac{\rho(t-T)+e^{-\rho(t-T)} - 1}{\rho^2} \bigg] \bigg\} \\& \quad + (1-\alpha) V  \bigg\{ \frac{1-e^{-\rho t}}{\rho} - \theta(t-T) \frac{1-e^{-\rho(t-T)}}{\rho}  \bigg\},
\end{split}\end{equation}
for $\beta=\lambda$.

\begin{figure}[!htb]
\centering
\includegraphics[width=8cm]{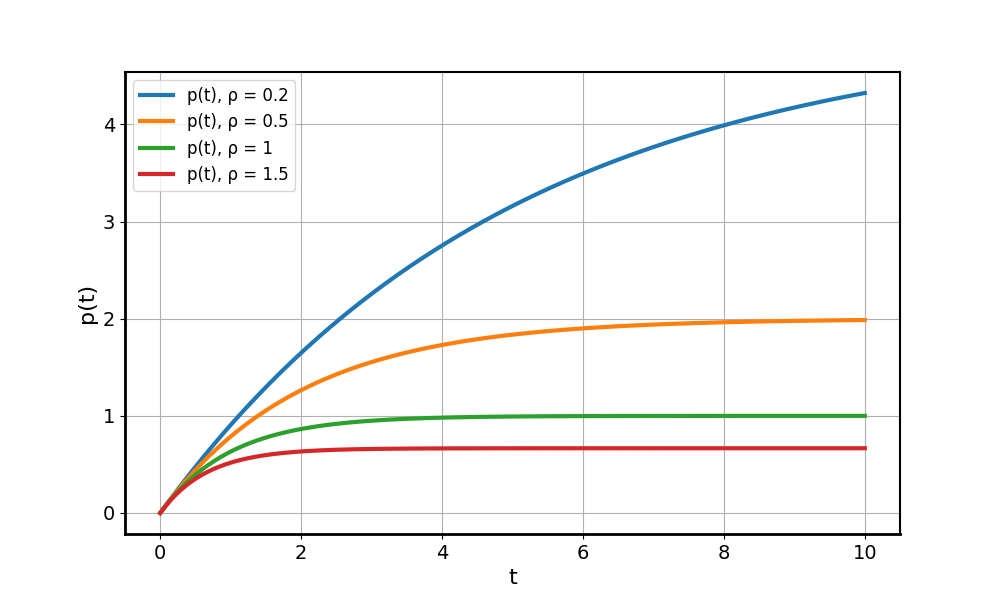}
\includegraphics[width=8cm]{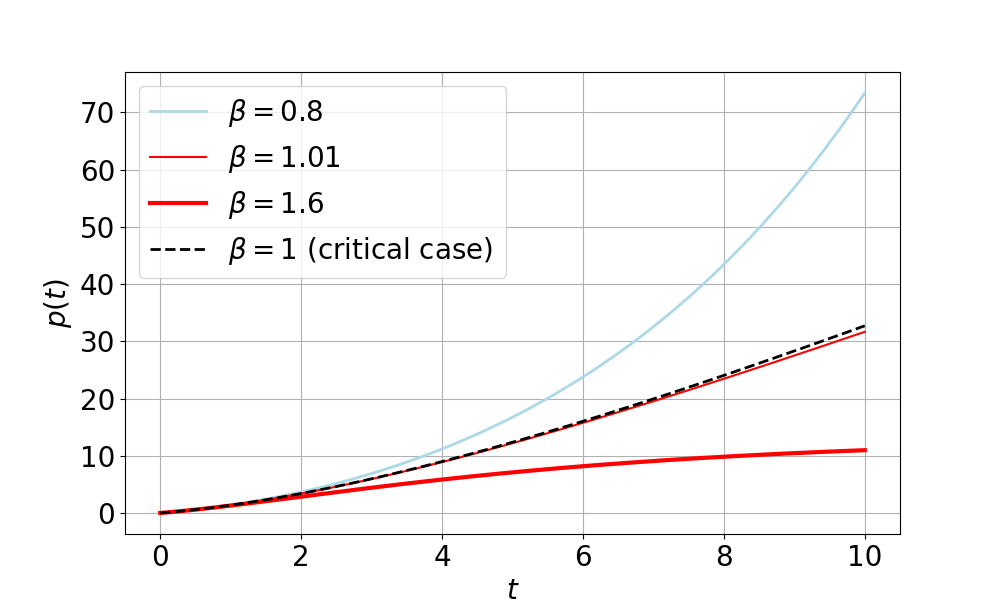}
\caption{The left panel shows the dynamics of prices for $\alpha=0$ for different values of the exponent $\rho$ for $\lambda=1, V=1$. The behaviour is always concave and convergent. The right panel, instead, shows the dynamics for $\alpha=1, \rho=0.2, \lambda=1, V=1$ ad different values of $\beta$. In this case, the regime depends on the sign of the difference $(\beta - \lambda)$.}
\label{fig:RNN_results}
\end{figure}

{\bf Price dynamics during the execution of the metaorder.} Let us first study the dynamics of prices during the execution of the metaorder, i.e. when $t<T$, which is
\begin{equation}\begin{split}\label{eq: PricesMOExecution}
    p(t<T) = \begin{cases}
         \alpha V \bigg[  \frac{\beta}{\beta - \lambda} \frac{1-e^{-\rho t}}{\rho} + \big( 1-\frac{\beta}{\beta-\lambda} \big) \frac{e^{(-\beta + \lambda)t  } - e^{-\rho t}}{\rho - \beta + \lambda} \bigg] + (1-\alpha) V   \frac{1-e^{-\rho t}}{\rho}, &  \text{for $\beta> \lambda$}\\
         \alpha V \big[ \frac{1-e^{-\rho t}}{\rho} + \beta ~ \frac{\rho t + e^{-\rho t}-1}{\rho^2} \big]  + (1-\alpha) V   \frac{1-e^{-\rho t}}{\rho}, & \text{for $\beta=\lambda$}.  
        \end{cases}
   \end{split}\end{equation}

Figure~\ref{fig:RNN_results} shows the behavior of prices for the two limiting cases, i.e. $\alpha=0$ and $\alpha = 1$. The left panel presents the behavior of the price for $\alpha=0$, which corresponds to the configuration in which the metaorder only influences the price dynamics without influencing market volumes. In this limit, there is no dependence on $\lambda$ and $\beta$, i.e. the parameter of the volumes kernel, but only on $\rho$, i.e. the parameter of the prices kernel. The price dynamics are concave for every $\rho>0$ and convergent to the value $p^*=V/\rho$ for sufficiently long metaorders, i.e.  $T\gg \rho^{-1}$.

The right panel illustrates the opposite limit, i.e., $\alpha = 1$, where price dynamics are influenced solely by metaorders through the volume dynamics. In this case, as with volumes, the reciprocal value of $\beta$ and the scale $\lambda$ become relevant. For $\beta > \lambda$, the price dynamics remains consistently concave. As we approach the critical condition $\beta = \lambda$, if $\alpha$ is appreciably different from zero, after the relaxation of price dynamics (which occurs at a rate determined by the inverse of $\rho$), the price tends to a \textit{linear} regime. Thus, if the system is near this regime, i.e. $\beta \gtrsim \lambda$, the concave convergence of prices is still guaranteed to occur, but on much longer timescales, depending on how close $\beta$ is to $\lambda$.

\begin{figure}[!htb]
\centering
\includegraphics[width=8cm]{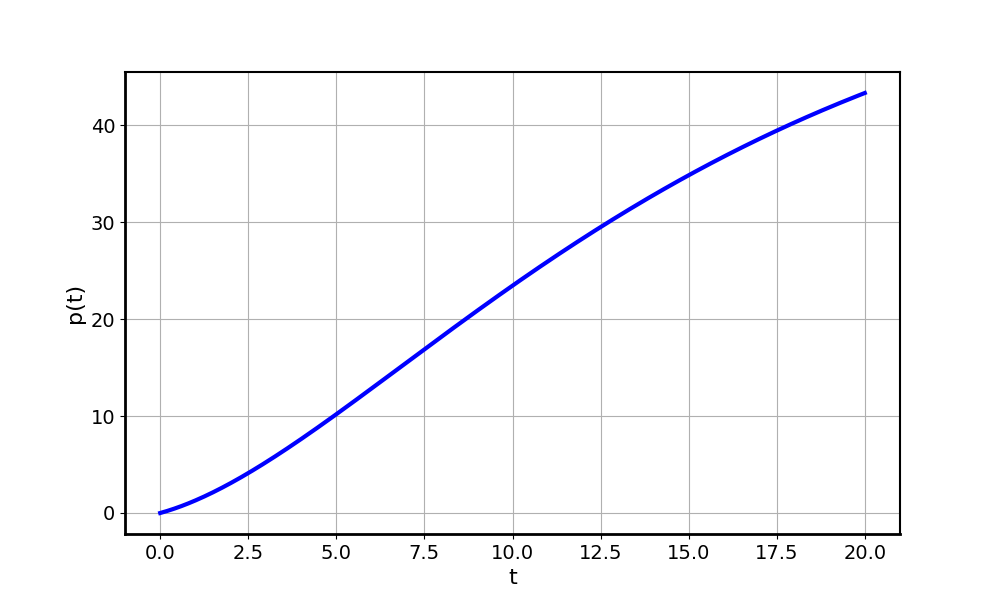}
\caption{Price dynamics during the execution of a metaorder under the modified TIM with  $\alpha=0.75, V=1, \lambda=1, \rho=0.1, \beta=1.15$. Note the change of concavity expected by the computations in Eq.~\eqref{eq: concavity}, which predicts a positive concavity for $t \ll \frac{1}{\beta-\lambda}$ and $t\ll \frac{1}{\rho}$. For our choice of the parameters, this corresponds to $t \ll 10$. As time grows, volume tends to a constant value and, thus, the curve changes concavity approaching to an asymptote.}
\label{fig:Prices_concavity}
\end{figure}

Let us now examine what happens at the beginning of the metaorder execution, i.e., for 
\begin{equation}\begin{split}
    t \ll \min \left(\frac{1}{\rho},\frac{1}{\beta - \lambda}\right).
\end{split}\end{equation}
By performing a Taylor expansion, we obtain
\begin{equation}\begin{split}\label{eq: concavity}
    P(t \sim 0) \simeq t V + t^2 V[ \alpha \lambda - \rho ], \quad \text{for $\beta > \lambda$}.
\end{split}\end{equation}
Thus, at small times, the price initially grows linearly, regardless of the value of $\alpha$. As $t$ increases, the quadratic term also becomes significant. If $\alpha = 0$, the behavior is always concave since $\rho > 0$. On the other hand, for $\alpha > 0$ and $\alpha > \rho / \lambda$, the price dynamics is initially convex, then it becomes concave as the price begins to converge to a fixed value. This change in concavity is illustrated in Fig.~\ref{fig:Prices_concavity}. In the opposite case, $\alpha < \rho / \lambda$, the price dynamics remains concave at all times.

Finally, for $\beta > \lambda$, price dynamics eventually becomes concave for sufficiently large times, converging to the value
\begin{equation}\begin{split}
    p(t) \rightarrow \alpha \frac{V}{\rho} \frac{\beta}{\beta - \lambda} + (1 - \alpha) \frac{V}{\rho}.
\end{split}\end{equation}

In contrast, as we approach the critical case $\beta = \lambda$, we tend to a configuration in which the linear term persists, as seen in Eq.~\eqref{eq: PricesMOExecution}, similar to the volume case. Thus, near criticality, since the volume during the metaorder only saturates for very large times, a similar behavior arises in prices, resulting in a quasi-linear price trajectory converging to a fixed value only on large timescales.

{\bf Price dynamics after the execution of the metaorder.} Let us now discuss what happens after the metaorder execution. Thus, let us switch on the terms proportional to the Heaviside function in Eq.s~\eqref{eq: PriceDynMain1}-\eqref{eq: PriceDynMain2}. For $t\geq T$, the expression for prices becomes 
\begin{equation}\begin{split}\label{eq: PricesAMO}
    P(t \geq T) & =  \alpha V \bigg\{ \frac{\beta}{\beta-\lambda} \frac{e^{-\rho(t-T)} - e^{-\rho t}}{\rho} + \bigg( 1 - \frac{\beta}{\beta - \lambda} \bigg) \bigg[ \frac{e^{-\rho(t-T)} - e^{-\rho t}}{\rho - \beta + \lambda} \\& + \frac{e^{(-\beta+\lambda)t}-e^{-(\beta-\lambda)(t-T)}}{\rho - \beta + \lambda} \bigg]\bigg\} +  (1-\alpha) V \frac{e^{-\rho(t-T)} - e^{-\rho t}}{\rho},
\end{split}\end{equation}
for $\beta > \lambda$, and 
\begin{equation}\begin{split}\label{eq: PricesAMOLin}
    P(t \geq T) & = \alpha V \bigg\{ \frac{e^{-\rho(t-T)} -  e^{-\rho t}}{\rho} + \beta ~ \frac{\rho T + e^{-\rho t} - e^{-\rho(t-T)}}{\rho^2} \bigg\} + (1-\alpha)V \frac{e^{-\rho(t-T)}- e^{-\rho t}}{\rho},
\end{split}\end{equation}
for $\beta=\lambda$. For sufficiently long metaorder, i.e. when  
\begin{equation}\begin{split}\label{eq: ConditionTaylorAfter1}
    T \gg \frac{1}{\rho},
\end{split}\end{equation}
we can approximate the expression of prices as
\begin{equation}\begin{split}
    P(t \geq T) \simeq 
    \begin{cases}
        \alpha V  \big[ \frac{\beta}{\beta-\lambda} \frac{e^{-\rho (t-T)}}{\rho} + \big( 1 - \frac{\beta}{\beta - \lambda} \big) \frac{e^{-\rho (t-T)} - e^{-(\beta-\lambda)(t-T)}}{\rho - \beta + \lambda} \big] + (1-\alpha) V \frac{e^{-\rho(t-T)}}{\rho}, & \text{for $\beta > \lambda$,}\\
        \alpha V  \big[ \frac{e^{-\rho(t-T)} }{\rho} + \beta~ \frac{\rho T - e^{-\rho(t-T)}}{\rho^2}\big] + (1-\alpha) V \frac{e^{-\rho(t-T)} }{\rho} & \text{for $\beta = \lambda$}.
    \end{cases}
\end{split}\end{equation}
In particular, it is interesting to study what happens immediately after the end of the metaorder, i.e. $t\sim T$. Thus, by taking the limit 
\begin{equation}\begin{split}\label{eq: ConditionTaylorAfter2}
    t-T \ll \min \left( \frac{1}{\rho}, \frac{1}{\beta - \lambda} \right),
\end{split}\end{equation}
we can approximate price dynamics as
\begin{equation}\begin{split}\label{eq: concavity2}
    P(t \sim T) \simeq 
    \begin{cases}
        \frac{1}{\rho} V  \big( \alpha \frac{\beta}{\beta-\lambda} + (1-\alpha) \big) - (t-T) V + (t-T)^2 (\rho - \alpha \lambda), & \text{for $\beta \neq \lambda$}, \\
        \frac{V}{\rho^2}\big( \alpha \beta ( T \rho - 1) + \rho \big) - (t-T) \frac{V}{\rho} \big[ \rho - \alpha \beta \big] + (t-T)^2 V \big[ \rho -  \alpha \beta \big], & \text{for $\beta = \lambda$}.
    \end{cases}
\end{split}\end{equation}
\begin{figure}[!htb]
\centering
\includegraphics[width=8cm]{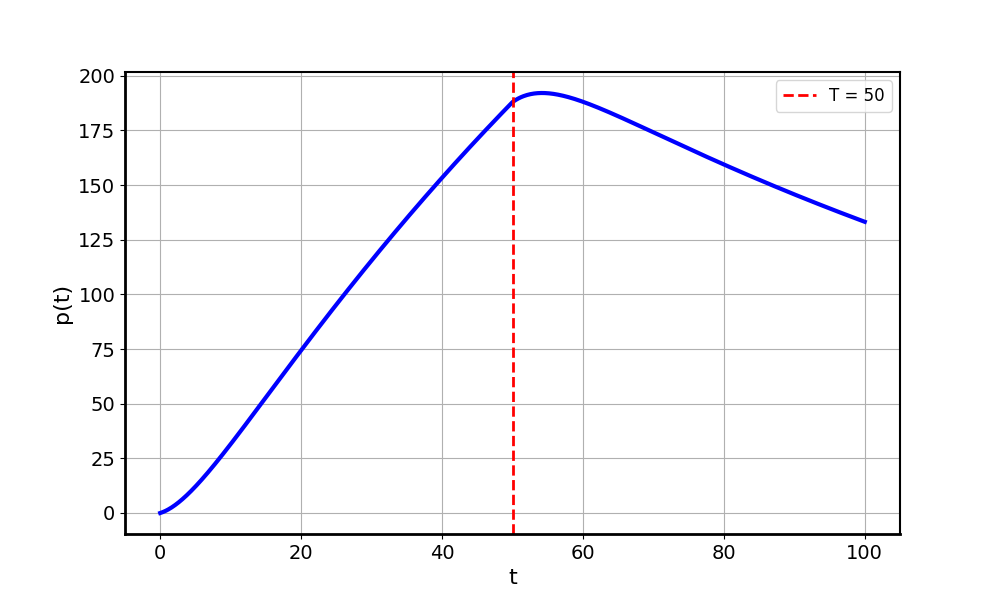}
\caption{Price dynamics during the execution of a metaorder of length $T=50$ under the modified TIM with $\alpha=0.8, V=1, \lambda=1, \rho=0.009,  \beta=1.2$
Note that after the end of the metaorder the price still increases for a time interval in which the conditions in Eq.s~\eqref{eq: ConditionTaylorAfter1} and \eqref{eq: ConditionTaylorAfter2} are satisfied. Then, the price dynamics always falls in a convex behavior. Note also the opposite concavity at the beginning of the metaorder and immediately after its end.}
\label{fig:Pt_Taylor_MO}
\end{figure}
We observe that the quadratic term has a very similar behaviour to the one obtained in Eq.~\eqref{eq: concavity} but with opposite signs. This means that, in general, the concavity immediately after the metaorder is always opposite to the one at the beginning of the metaorder described in Eq.~\eqref{eq: concavity}. 
Thus, if $ \alpha > \rho/ \lambda$, the price behaviour after the metaorder is concave for a certain period of time, established by the conditions in Eq.s~\eqref{eq: ConditionTaylorAfter1} and \eqref{eq: ConditionTaylorAfter2}, and the price continues to increase also after the end of the metaorder. An example of this is shown in Fig~\ref{fig:Pt_Taylor_MO}. This is the ``inertial" effect we observed in the previous Sections when simulating the price trajectory according to empirically estimated linear or nonlinear models. Differently, when $ \alpha < \rho/ \lambda$ the price behaviour after the execution is always convex resulting in a immediate and more rapid fall of the price. 

In the opposite limit, i.e. $(t-T) \gg 1$, the price tends to an asymptotic value. When $\beta>\lambda$ this limit is zero, independently from $\alpha$, i.e. 
\begin{equation}\begin{split}\label{eq: AsymMO}
    \lim_{t \rightarrow +\infty}p_{\beta > \lambda} (t) = 0.
\end{split}\end{equation}
Thus the price impact is completely transient as in the TIM.
On the contrary, when $\beta = \lambda$ it is  
\begin{equation}\begin{split}\label{eq: AsymMOLin}
    \lim_{t \rightarrow +\infty}p_{\beta=\lambda}(t) = \frac{\alpha V \beta T}{\rho},
\end{split}\end{equation}
and the long term limit is  different from zero unless $\alpha=0$. In this critical case and under the assumption that the metaorder triggers some volume from the market ($\alpha\ne 0$) one observes a permanent impact of the metaorder.

We found that prices can converge to a nonzero asymptote, leaving a permanent impact on the market, even with exponential kernels. This occurs under the critical condition \(\beta = \lambda\). However, as in the case of volumes, observing a long-memory impact does not strictly require \(\beta\) to be exactly equal to \(\lambda\). In fact, when \(\beta \gtrsim \lambda\), the metaorder information may still decay very slowly to zero over timescales that become increasingly long as \(\beta\) approaches \(\lambda\). 

Near the critical regime, for \(\alpha\) sufficiently close to 1, a linear growth dominates during the execution of the metaorder, when the exponential terms relax to zero. Moreover, as shown in Eq.~\eqref{eq: AsymMOLin}, in the critical limit, the price converges to a nonzero asymptote proportional to \(\alpha\). Therefore, once the model parameters are fixed, the value of \(\alpha\) significantly influences how much the price relaxes after the metaorder execution. 

In particular, if \(\alpha > \rho / \lambda\), the price exhibits a concave behavior after the metaorder ends. This behavior is discussed in greater detail in the following paragraphs.

\subsubsection{Two different relaxations after metaorder execution}

In the previous discussion it emerged that, even with short range kernels, it is possible to observe long impact effects if the system is close to criticality. In the price equation for \(\beta \neq \lambda\) in Eq.~\eqref{eq: PricesAMO}, two relaxation processes occur, potentially on different time scales. The first one is governed by the scale \(\rho\), which characterizes the price kernel. The second relaxation is governed by the difference \((\beta - \lambda)\). Notice that this is different from \(\beta\), which describes the relaxation scale of the volume kernel. Therefore, even if \(\beta\) and \(\rho\) are similar, the relaxation due to these two kernels could occur on significantly different time scales, depending on the difference between \(\beta\) and the characteristic frequency scale \(\lambda\). As a result, the dynamics of volumes could retain a longer memory, which may never fully relax in the critical limit \(\beta = \lambda\). In this case, volumes maintain a permanent memory, as shown in Eq.~\eqref{eq: VolDynamicsMainRelax}, while the price kernel exponentially loses metaorder information.

\begin{figure}[!htb]
\centering
\includegraphics[width=8cm]{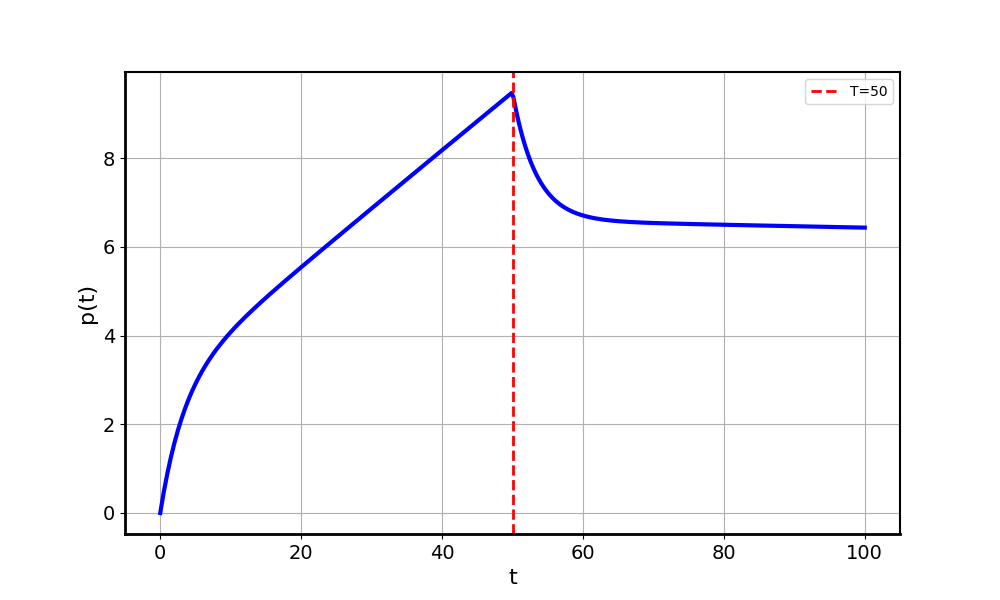}
\caption{Price dynamics during the execution of a metaorder of length $T=50$ under the modified TIM  with $\beta = 0.401\sim \lambda = 0.4$, $\rho=0.3$, $V=1$, $\alpha=0.1$. After a phase of concave growth followed by a linear one, two stages of information relaxation can be observed for $t > T$. The first relaxation, due to price dynamics, occurs when $(t-T) \sim 1/\rho$. The second, due to volume dynamics, is much slower, as volumes retain metaorder information, converging to zero only at much longer timescales.}
\label{fig:Pt_MO_Relax}
\end{figure}

Thus, in an exponential model, our framework relaxes metaorder information in price in two stages, regulated by two different frequency scales. The first relaxation occurs immediately after the end of the metaorder and is governed by the price kernel parameter \(\rho\). Since \(\rho > 0\), in Eq.~\eqref{eq: PricesAMO}, when $(t-T) \sim 1/\rho$, the exponential terms involving \(\rho\) decay to zero, and the price simplifies to
\begin{equation}\begin{split}\label{eq: FirstRelax}
    P(t-T \sim 1/\rho) & =  \alpha V \bigg( 1 - \frac{\beta}{\beta - \lambda} \bigg)  \frac{e^{(-\beta+\lambda)t}-e^{-(\beta-\lambda)(t-T)}}{\rho - \beta + \lambda},
\end{split}\end{equation}
which, in the case \(\beta = \lambda\), reduces directly to Eq.~\eqref{eq: AsymMOLin}. The remaining difference between this price and the initial price, which is zero in our setup, represents the information still retained by the volumes.

At this point, two scenarios are possible. In the first, \(\beta\) is significantly larger than \(\lambda\). In this case, the price continues to relax and eventually converges exponentially  to its initial value. In the second scenario, the volume retains the metaorder information. If \(\beta\) is sufficiently close to \(\lambda\), volumes ``stabilize" at a non-zero value (i.e., different from the initial value), and the same occurs for prices. This stabilization of prices after the first relaxation could be permanent for \(\beta = \lambda\) (resulting in a permanent impact) or quasi-stable for \(\beta \gtrsim \lambda\) (resulting in a transient, but long lived, impact). 

In Fig.~\ref{fig:Pt_MO_Relax}, we present an example of possible price dynamics generated by our model for \(\beta \sim \lambda\) and \(\alpha = 0.1\). The figure illustrates a phase of concave growth during the metaorder execution, followed by a linear growth. Immediately after the execution, for \((t-T) \sim 1/\rho\), a convex relaxation occurs due to the influence of the exponential term with \(\rho\). This concave growth and convex decay are ensured by choosing \(\alpha < \rho/\beta\), as suggested by Eqs.~\eqref{eq: concavity} and \eqref{eq: concavity2}. Finally, the price exhibits a slower relaxation phase, influenced by the exponentials involving \((\beta - \lambda)\).

In the framework of this model with exponential kernels, metaorders should always influence volume dynamics to maintain impact in the market, i.e., $\alpha \neq 0$, because if they do not, the only relevant timescale would be the price frequency scale $\rho$. In this case, after the metaorder, the price would always revert to its initial value exponentially, without retaining any memory of the metaorder. Additionally, it is crucial for the system to be near the critical condition for volumes, i.e., $\beta \sim \lambda$, because if this condition is not satisfied, the price would still exponentially converge to zero even when $\alpha \neq 0$. Thus, if the system is close to the critical condition, an impact can be sustained in the market for a long period, even in a scenario governed by exponential dynamics.

\subsubsection{An explanation of the linear growth and the partial decrease of prices}

Recently, it has been observed in the literature that when models are fitted on public data and then used to generate expected metaorder dynamics, they often fail~\cite{cont2023limit, elomari2024microstructure}. In particular, during simulated metaorders, it is frequently observed that prices exhibit linear growth instead of the expected concave behavior, followed by only a partial convex decrease, with stabilization occurring near the peak.

This is in contrast with what is typically observed in real metaorder data, where prices display concave growth during the metaorder, followed by a noticeable convex decrease that converges to a semi-stable point. This dichotomy between empirical results and the predictions of linear models suggests that linear models may not adequately capture the market's response to metaorder executions. However, as discussed earlier in this paper, we have also shown that highly nonlinear models, such as neural networks, could similarly fail to capture these market dynamics. This raises the possibility that we may be misapplying both linear and nonlinear models when simulating metaorders.

Some insights into why this occurs can be drawn from the model discussed earlier. In the previous section, we noted that when the system is close to the critical point, i.e., $\beta \gtrsim \lambda$, the price trajectory exhibits long memory effects even in an exponential world due to a long impact in volumes. However, it was shown that unexpected behaviours could occur when the metaorder contribution is included only in the volumes dynamics (\(\alpha=1\)) in these conditions. This suggests that when the used model is close to criticality, further attention is needed in how the metaorder information is incorporated into the simulation. And this seems to be the case. Indeed, when we fit linear models on financial public data we often observe conditions that are close to criticality.

We discussed that in the limit $\beta \gtrsim \lambda$, a linear regime replaces the concave one for sufficiently large $t$, i.e., $t \sim 1/\rho$. This can be seen from Eq.~\eqref{eq: PricesMOExecution}. Additionally, it is evident that the closer $\alpha$ is to 1, the more significant the linear term becomes. Therefore, if the metaorder is only incorporated into the volume dynamics, i.e., $\alpha = 1$, and the system is near criticality, the price growth will be predominantly linear. Furthermore, note that $\alpha$ also appears in Eq.~\eqref{eq: AsymMOLin} and Eq.~\eqref{eq: FirstRelax}. Thus, the larger $\alpha$ is, the higher the ``first asymptote" will be.

\begin{figure}[!htb]
\centering
\includegraphics[width=8cm]{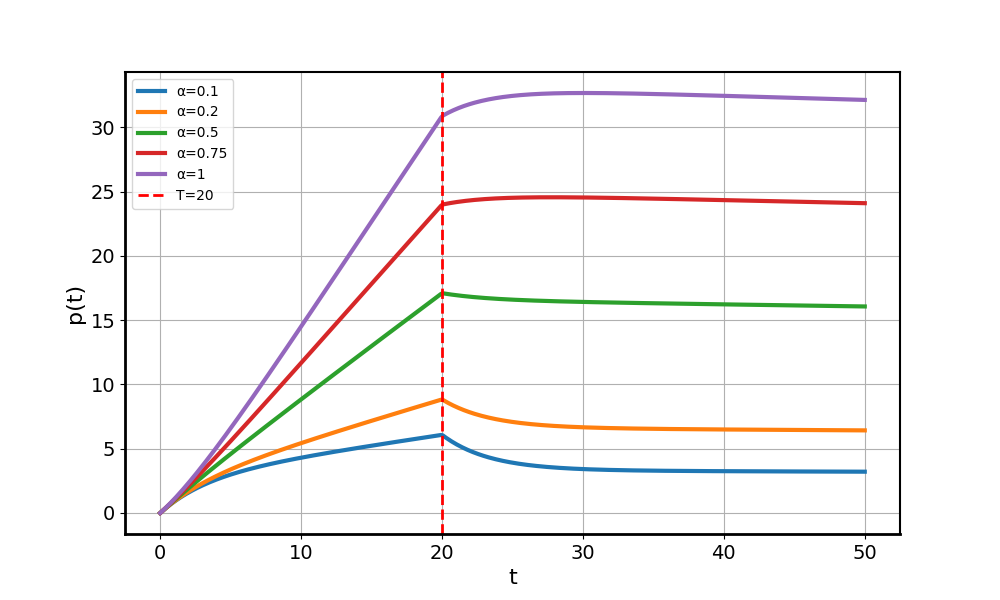}
\caption{Price dynamics for different values of $\alpha$ and for values of $\beta$ and $\lambda$ close to the critical limit, i.e. $\beta = 0.501\simeq \lambda=0.5$, and with $\rho=0.3$, $T=20$, $V=1$.}
\label{fig:Linear_Expl}
\end{figure}
Now, considering the critical limit in Eq.~\eqref{eq: PricesAMOLin}, the peak value of the price at the end of the metaorder is given by:
\begin{equation}\begin{split}
    p_{\beta=\lambda}(T) = \frac{\alpha V \beta T}{\rho} + \frac{\alpha V \beta (e^{-\rho T} - 1)}{\rho^2} + V \frac{1-e^{-\rho T}}{\rho} \simeq p_{\beta = \lambda}(t \rightarrow \infty) + \frac{V}{\rho^2}(\rho - \alpha \beta),
\end{split}\end{equation}
where we used Eq.~\eqref{eq: AsymMOLin} and approximated the exponential term to zero for sufficiently long metaorders. This gives us:
\begin{equation}\begin{split}
    p_{\beta=\lambda}(T) - p_{\beta = \lambda}(t \rightarrow \infty) \simeq \frac{V}{\rho^2}(\rho - \alpha \beta),
\end{split}\end{equation}
which indicates that the closer \(\rho\) and \(\alpha \beta\) are, the smaller the decrease during the first relaxation will be (see Fig.~\ref{fig:Linear_Expl}). Furthermore, if \(\alpha > \rho/\beta\), the first asymptote will be larger than the peak, meaning that the price continues to increase. This is consistent with our earlier discussion around Fig.~\ref{fig:Pt_Taylor_MO} and reinforces the condition \(\alpha < \rho/\beta\).

In Fig.~\ref{fig:Linear_Expl}, we illustrate the behavior of the price near the critical point for different values of \(\alpha\). As \(\alpha\) increases, the price behavior after the metaorder deviates significantly from the expected convex decrease. This indicates that fully incorporating the metaorder into the volume dynamics near critical conditions can result in unexpected price behaviors. Smaller values of \(\alpha\), on the other hand, help mitigate the long-memory effects of criticality, as the autoregressive contribution of volumes on prices becomes suppressed. In the following Section, we extend these observations to the hyperbolic case.

\subsection{Discrete time model}

The closed form expressions in the previous Section are obtained under the assumption of exponential kernels. Unfortunately, the analytical computations become much harder for hyperbolic kernels. To test the robustness of our results to the choice of kernel, we consider the discrete version of the model. In Appendix~\ref{Appendix_Discretisation}, we report in details the discretisation of model~\eqref{eq: VolterraEqs} which reads 
\begin{equation}\begin{split}\label{eq: DiscreteModel}
    & p_t = \sum_{i=1}^t g_{i} [v_{t-i} + (1-\alpha) V \theta(T-(t-i)) ], \\&
    v_t = \alpha V  \theta(T-t) + \lambda \sum_{i=1}^t d_{i} v_{t-i},
\end{split}\end{equation}
where \(\theta(T-t)\) is the discrete Heaviside function.  The second equation is an autoregressive model and the stationarity condition reads 
$$
\lambda\sum_{i=1}^\infty d_i<1
$$
while the critical case corresponds to the case when the sum is exactly equal to one.

\begin{figure}[!htb]
\centering
\includegraphics[width=8cm]{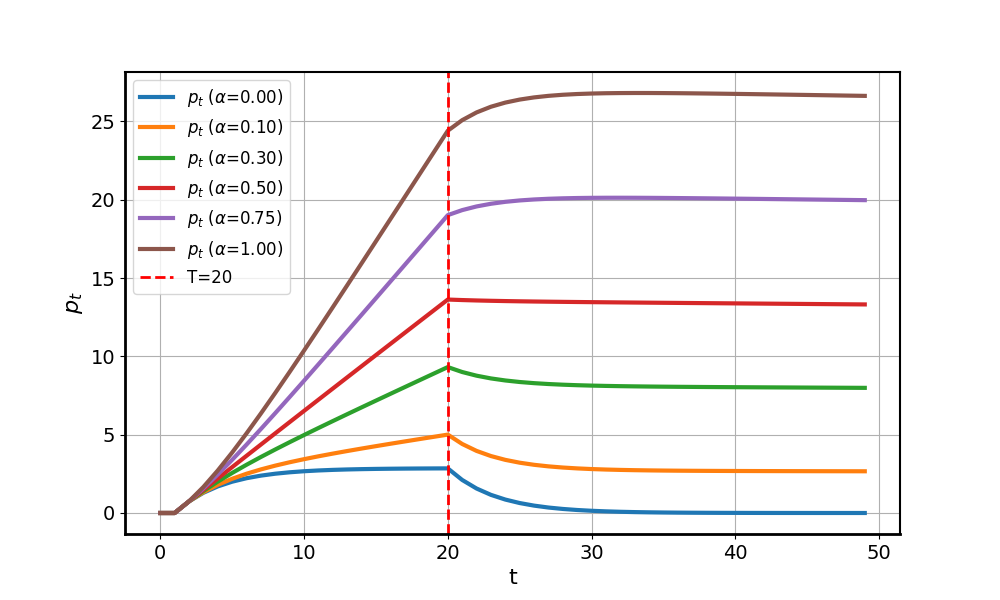}
\caption{Price dynamics of the discrete model in the exponential case for $\beta= 0.7$, $\rho = 0.3$, $T=20$, $\lambda=1$, $V=1$ for different values of $\alpha$. The values have been chosen such that we are close to the critical condition.}
\label{fig:Calcolo_Pt_Discrete_Exp}
\end{figure}

{\bf Exponential kernels.} We first show that the results of the discrete model with exponential kernels are in line with those obtained analytically in the continuous case. The coefficients of the model are  
\begin{equation}\begin{split}
    d_i = e^{-\beta i}, \quad g_i = e^{-\rho i},
\end{split}\end{equation}
and the critical condition is $\lambda (1+e^{\beta})^{-1}=1$.
Figure~\ref{fig:Calcolo_Pt_Discrete_Exp}, shows the results for values of $\beta$ and $\lambda$ such that we are close to such critical condition. Similarly to the continuous case, when we are close to criticality, as \(\alpha\) approaches 1, the price trajectory displays unexpected behaviors, such as linear growth and an extended memory effect.

\begin{figure}[!htb]
\centering
\includegraphics[width=8cm]{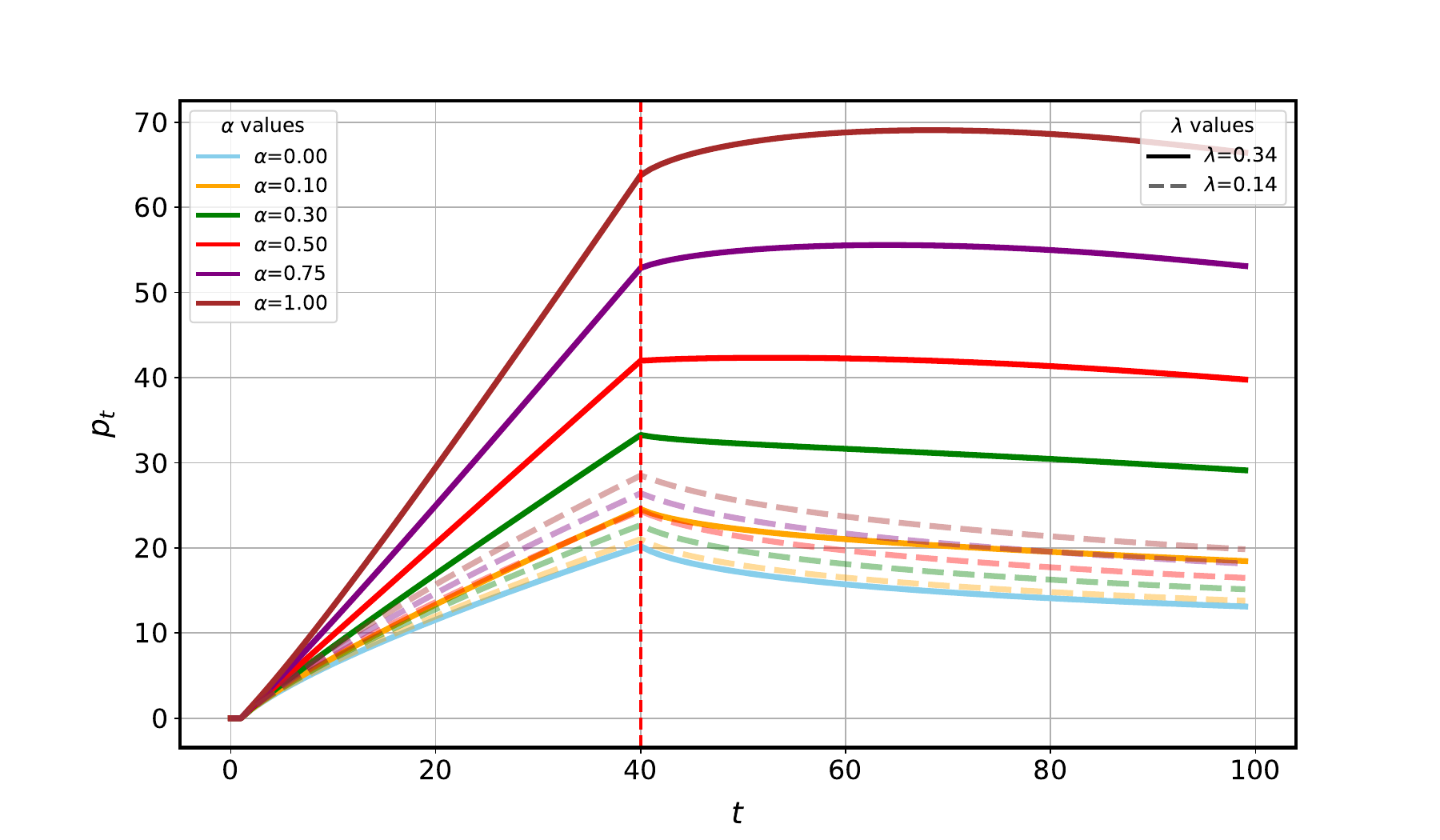}
\includegraphics[width=8cm]{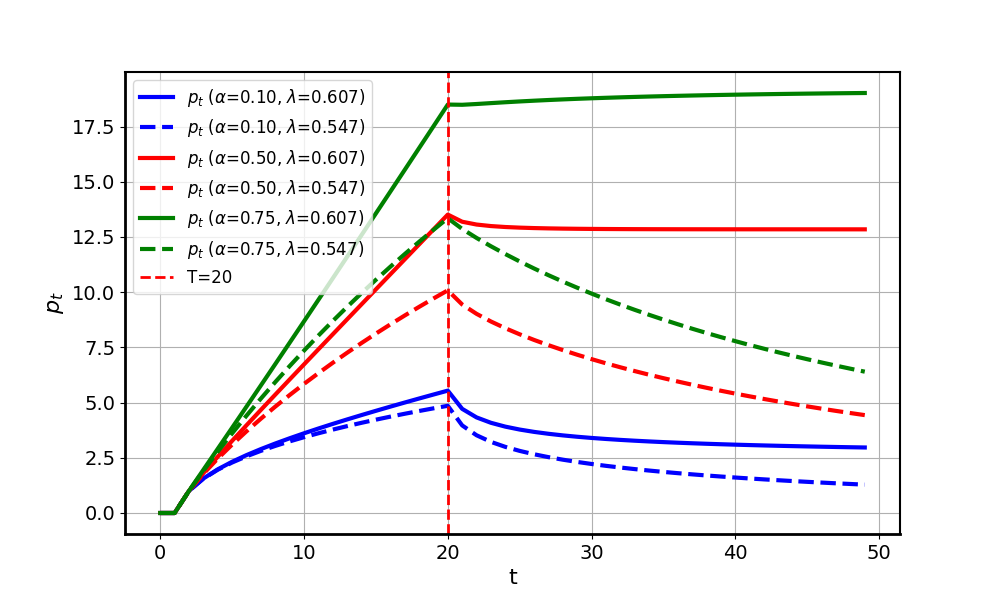}
\caption{Left panel: price dynamics for the discrete model in the power-law case for $\eta = 1.5$, $\delta= 0.25$, $T=40$, $V=1$. The solid lines represent the case $\lambda = 0.34$ where the system is close to criticality ($\lambda \sum d_i \sim 0.8$) while the dashed lines represent the case $\lambda = 0.14$, where we are far from criticality. Right panel: price dynamics for the discrete model in the hyperbolic case for $\eta=2$, $\delta=1$, $T=20$, $V=1$ for two values of $\lambda$. The figure shows that in the case of faster decreasing exponent criticality becomes important for memory effects.}
\label{fig:Calcolo_Pt_Discrete}
\end{figure}

\newpage
{\bf Power-law kernels.} 

To obtain numerical results when kernels decay asymptotically as a power-law function, we choose 
\begin{equation}\begin{split}
    d_i = \frac{1}{i^\eta}, \quad g_i = \frac{1}{i^\delta}.
\end{split}\end{equation}
The criticality condition is in this case $\lambda \zeta(\eta)=1$, where $\zeta$ is the Riemann zeta function. To fix realistic parameters, we remind that Ref. \cite{Bouchaud2003FluctuationsAR} shows that the price is diffusive if the relation \(\delta = \frac{1-\gamma}{2}\), where \(\gamma\) is the exponent of the power law describing the autocorrelation of volumes, is satisfied. Empirical data show that reasonable values are  \(\gamma \sim 0.5\), thus \(\delta \sim 0.25\) and \(\eta=1+\gamma \sim 1.5\). This choice puts an upper bound on \(\lambda\) to satisfy stationarity.

The left panel of Fig.~\ref{fig:Calcolo_Pt_Discrete} shows the price dynamics for different values of $\alpha$.

When we are close to criticality (solid lines) the qualitative behavior of price is similar to the one obtained for the exponential kernel. When $\alpha$ is close to $1$ the price increases non-concavely and after the end of the metaorder the reversion of price is not appreciable on small time scales. On the contrary, a smaller value of \(\alpha\) results in a more typical concave growth and convex decrease, as observed empirically. This occurs because volume's critical contributions are suppressed by a small value of \(\alpha\). Consequently, most of the metaorder contribution would directly influence the price dynamics without passing from volumes. As expected, when we are far from criticality (dashed lines), these effects tend to be suppressed for every $\alpha$, as shown by the dashed lines in the left panel of Fig.~\ref{fig:Calcolo_Pt_Discrete}. The right panel of Fig.~\ref{fig:Calcolo_Pt_Discrete} makes a comparison of the price trajectory for different values of $\alpha$ and two values of $\lambda$, the largest one putting the system close to criticality. In this case, faster decreasing exponents make the presence of criticality more important to maintain a longer impact. 

At this point, one could observe that criticality and long-range kernels appear to play a similar role in maintaining long impact. Indeed, when prices are described by long-range kernels, a critical value of \(\lambda\) is not required to sustain long-term impact on prices. This is because, even if volumes relax faster (e.g., as \(t^{-1.5}\), shown in the left panel of Fig.~\eqref{fig:Calcolo_Pt_Discrete}), prices can still exhibit long-term impact due to the small value of the kernel exponent. However, criticality becomes significantly more important when short-range kernel are considered, as in the exponential case. In the right panel of Fig.~\ref{fig:Calcolo_Pt_Discrete}, we illustrate that if \(\lambda\) deviates from its critical value and the kernel is not sufficiently long-range, the impact relaxes appreciably to zero.

In conclusion, we observe that, when we are close to the critical condition, inserting $\alpha=1$ triggers large effects in volume dynamics that, in turn, lead to the similar effects in prices, which do not relax as expected. Thus, if calibrating our model on data leads to be close to criticality, it is necessary to pay attention on how the metaorder is inserted in the volume dynamics. Indeed, decreasing the value of $\alpha$ suppresses this critical contribution making the concavity and the relaxation more evident. Instead, if we insert the metaorder contribution entirely in volumes, we would overestimate the effect of volumes on price dynamics. Furthermore, from our results, for long memory kernels of prices the critical condition on volumes seems to be not important to maintain an impact. Indeed, even if volumes relax faster, prices relax slower. The critical condition becomes, however, much more relevant for shorter tailed kernels in which volume dynamics plays a fundamental role in maintaining long impact in the market.

\section{Discussions and Conclusions}\label{sec:conclusions}

In this paper, we have shown that models of price and trade calibrated on public market data typically produce price trajectories during and after the execution of a metaorder that differ significantly from those observed when proprietary data of real metaorders are used. In particular, the price increases almost linearly during the execution and reverts only marginally after its completion, stabilizing at a finite value and leaving a permanent impact. In our simulations, this behaviour emerges when the autoregressive term in the volume equation is included, when the number of lags is large, since the cumulative effect of its coefficients becomes strong enough to drive the system close to instability. This occurs in particular when the child order affects the dynamics of the order flow, which in turn is incorporated into the price formation process. Although this coupling could, in principle, be avoided by introducing the child order’s contribution directly into the price equation, the effect becomes especially relevant in order book models, where the influence of the child order on the order flow is inherently unavoidable. An example of this effect can be observed in~\cite{elomari2024microstructure}, where the model describing the LOB lives close to instability and simulating the price impact the authors are not able to obtain the expected stylized behaviour of price. The results contained in this paper suggest that somehow the effects of child orders on volumes should be suppressed, especially when the system lives near instability. In fact, in that case, the effects of the child orders on volumes are largely persistent.

Consequently, from a structural point of view, our conclusions can be explained by considering that most of the autocorrelation of order flow is due to the simultaneous execution of several metaorders, as suggested by Lillo, Mike, and Farmer in \cite{lillo2005theory}. In their stylized model, the addition of a new metaorder does not trigger any additional order flow although the autocorrelation of order flow is very persistent. In other words, despite the statistical models correctly capture the correlations, they miss the causality behind them. This explanation can be further explored in light of empirical evidences on how the market reacts to the introduction of order flow \cite{Tóth01072012,doi:10.1142/S2382626618500028}.

On the other hand, from a statistical perspective, the fact that the estimated linear model operates near critical conditions also plays a relevant role, as the effects of child orders on the order flow, if not suppressed, generate large and persistent impacts on prices. The observation that the parameters of certain models calibrated on financial data tend to lie close to regions of criticality or instability is well established in the literature. This may result from the model’s inability to fully capture all the relevant features of the real system. Although the literature has not yet reached a definitive conclusion on the underlying causes of this phenomenon, it remains an empirical fact. Consequently, we must take it into account during model calibration, understand its implications, and develop suitable strategies to handle it. Therefore, the aim of this paper is \emph{not} to explain why this occurs, but rather to provide insights into how to deal with a price and trade model that, once calibrated, operates close to criticality in order to reproduce the stylized facts of metaorders.
The analytical and numerical examples proposed in this paper indicate, in fact, that close to criticality the price increases almost linearly and remains essentially constant after the end of the metaorder, as seen with linear and non-linear models.

We proposed a modified TIM model that depends on a parameter \(\alpha\), which scales the child order volume contribution in the price equation.
From the structural point of view, \(\alpha\) describes  the part of the trading volume of a child order, which triggers additional order flow from the market. The extreme case of the LMF model corresponds to $\alpha=0$, while the purely statistical model (e.g. a VAR) corresponds to $\alpha=1$. In general, one expects that in real data the value of this parameter might be intermediate between these extremes. With a single free parameter $\alpha$ modulating this effect, we show that price trajectories of our model are much closer to the one observed empirically. Interestingly, at the critical condition, market impact after the end of the metaorder becomes permanent also with short range kernels. Even if it is hard to justify why markets should live at this critical point, we show that close to criticality impact can decay extremely slowly.
From the statistical point of view, \(\alpha\) balances the long-term impact effects induced by criticality. Indeed, in the continuous model with exponential kernels, we observe that, in the presence of criticality, volumes exhibit a linear trend during the execution of the metaorder and subsequently become constant after execution. This behavior is inherited by prices, which, after the relaxation of the kernels, exhibit a persistent term, initially linear and then constant. The role of the parameter $\alpha$, therefore, is to reduce the contribution that prices inherit from volumes in their dynamics and restore the concave and then convex dynamics. A direct analytical resolution in the power law case becomes more challenging. For this reason, we performed simulations of the model in the discrete case, which qualitatively shows the same behaviors observed in the exponential case near the critical point. A more in-depth investigation in this direction is deferred to future works.

Although the models provide interesting insights on market impact, several questions remain open. First of all, it would be interesting to empirically estimate $\alpha$. While this should be quite straightforward with real metaorder data, it is less clear how it can be estimated with public market data. Moreover, our setting (as typical in great part of the market impact literature) considers only child orders executed as market orders. A natural question is therefore if
in a more general setting, where limit orders and cancellations are considered,  it is possible to disentangle the part of (global) order flow which reacts to child orders (for example stimulated refill) from that which is insensitive (because part of trading schedule of other metaorders). This question is particularly relevant if, instead of considering price and trade models, one is interested in full limit order book models. However, we expect that a pure statistical model might not be able to reproduce the price trajectory during and after a metaorder execution and, most likely, the associated cost.

\section*{Acknowledgements} 
Authors acknowledge Michael Benzaquen and Salma Elomari for useful discussions. 

This publication has received funding from the  European Union – NextGenerationEU – PNRR – Project: “SoBigData.it – Strengthening the Italian RI for Social Mining and Big Data Analytics” – Prot. IR0000013.

The research was carried out also with the contribution of research groups linked to the Horizon 2020 Program under the scheme “INFRAIA-01-2018-2019 – Integrating Activities for Advanced Communities”, Grant Agreement n.871042, ``SoBigData++: European Integrated Infrastructure for Social Mining and Big Data Analytics”, to the HORIZON Europe Program under the scheme ``INFRA-2021-DEV-02- SoBigData RI Preparatory Phase Project" Grant Agreement n. 101079043 — SoBigData RI PPP and to the PRIN project {\it Dynamic models for a fast changing world: An observation driven approach to time-varying
parameters} (Grant Agreement No. 20205J2WZ4).

\newpage
\bibliographystyle{unsrt}
\bibliography{Bibliografia}

\appendix
\section*{SUPPLEMENTARY MATERIAL}

\section{Generalised Impulse Response Analysis for metaorders}\label{GIRA}
To derive the expression of expected price dynamics during a metaorder execution in the H and TIM models, we assume that for $t = 1, \dots, T$ on the top of the order flow $v_t$ is imposed an order execution with volume $\delta_v$ for each transaction. Moreover if $\tau$ is the average inter-trade time (measured in seconds), the trading speed is $T \delta_v / (T \tau) = \delta_v / \tau$. 
All the above models can be written in companion form $z_t=\Gamma z_{t-1}+\epsilon_t$, where $\Gamma$ has dimension $2p \times 2p$. We then define a $2p \times 1$ vector $e_2$. For the H and TIM models, it is simply defined as $e_2:=(b_0,1,0,\dots,0)'$.

Taking the unconditional expectation, we consider the evolution of the models written in the companion form when they are also subject to the shocks due to each trade of size $\delta_v$ due to the metaorder. Indicating with $\tilde{z}_t := \mathbb{E}[z_t]$, its evolution is given by
\begin{equation}\begin{split}\label{svarCompC}
    \tilde{z}_t = \Gamma \tilde{z}_{t-1} + \delta_v e_2.
\end{split}\end{equation}
 Iterating Eq. \eqref{svarCompC} and using the fact that $\sum_{j=0}^{k-1} \Gamma^j = (I_{2p}-\Gamma)^{-1}(I_{2p}-\Gamma^k)$, we have that:

\begin{align}\label{startrk}
         \tilde{z}_1 &= \Gamma \tilde{z}_0 + \delta_v e_2 \\ 
        \tilde{z}_2 &= \Gamma \tilde{z}_1 + \delta_v e_2 = \Gamma^2 \tilde{z}_0 + \delta_v \Gamma e_2 + \delta_v e_2 \\ \nonumber
        \vdots \\ 
    \tilde{z}_k &= \Gamma^k \tilde{z}_0 + \sum_{j=0}^{k-1} \delta_v \Gamma^j e_2 = \Gamma^k \tilde{z}_0 + \delta_v (I_{2p}-\Gamma)^{-1}(I_{2p}-\Gamma^k)e_2,
\end{align}
where $I_{2p}$ is the $2p$ x $2p$ identity matrix.
 
Now we average over the possible histories of the unperturbed $\tilde{z}_t$ with $t \leq 0$, i.e. $\tilde{z}_0 = (0,\ 0, \dots, 0)'$ and we get the definition of our \textit{nonstandard} IRF

\begin{equation}\begin{split}\label{endrk}
    \tilde{z}_k =  \delta_v (I_{2p}-\Gamma)^{-1}(I_{2p}-\Gamma^k)e_2.
\end{split}\end{equation}

Eq. \eqref{endrk} follows the same logic as the standard IRF of Eq. \eqref{irf}, with the difference that the ``shock" $\delta_v$ is induced in the system at each trade $k$.

To get $\tilde \Delta p_k:={\mathbb E}[\Delta p_k]$, it is sufficient to premultiply Eq. \eqref{endrk} by the selection $2p \times 1$ vector  $e_1 := (1,0,\cdots,0)'$, and we get the following expression
\begin{equation}\begin{split}\label{rk}
    \Delta \tilde{p}_k =  \delta_v \Big[e_1'(I_{2p}-\Gamma)^{-1}(I_{2p}-\Gamma^k)e_2\Big].
\end{split}\end{equation}

Let $\Delta p_t$ be given by Eq.~\eqref{rk} and let the maximum eigenvalue of $\Gamma$ in absolute value lower than 1. From basic algebra, we have that $\sum_{j=1}^{k} \Gamma^j = \Gamma (\tilde{I}_{2p}-\Gamma)^{-1} (\tilde{I}_{2p}-\Gamma^k)$. Summing $\Delta \tilde{p}_j$ over $j$, after some basic computations, we get
\begin{align}
    \tilde{p}_k & = p_0 + \sum_{j=1}^{k}\Delta \tilde{p}_j \\& = p_0 + \sum_{j=1}^{k} \delta_v \Big[e_1'(\tilde{I}_{2p}-\Gamma)^{-1}(\tilde{I}_{2p}-\Gamma^j)e_2\Big] \\ 
        &= p_0 + \delta_v e_1'  (\tilde{I}_{2p}-\Gamma)^{-1}\Big[\sum_{j=1}^{k}(\tilde{I}_{2p}-\Gamma^j)\Big]e_2 \label{pkdim} .
\end{align}
Finally, expressing the expected midprice $\tilde{p}_k$ as the cumulative sum of the average price increments, we obtain
\begin{equation}\begin{split}\label{pkApp}
    \tilde{p}_k  & = p_0 + \sum_{j=1}^{k}\Delta \tilde{p}_j \\& =  p_0 + \delta_v \Big[ e_1' (I_{2p} - \Gamma)^{-1}\\&\times (k I_{2p}  - \Gamma (I_{2p}-\Gamma)^{-1}(I_{2p}-\Gamma^k))e_2\Big].
\end{split}\end{equation}

 Eq.~\eqref{pk} and Eq.~\eqref{pTkCorp} describe, respectively, the dynamics of the price during and after the execution of the metaorder in our setting. Despite the existing debate about the functional fit of the price trajectory,\footnote{For instance, \cite{almgren2005direct,toth2011anomalous,mastromatteo2014agent,brokmann2015slow} argue that the price trajectory during the execution obeys the "square-root law", while \cite{zarinelli2015beyond} found that a logarithmic functional form fits better.} the literature agrees on the fact that the price trajectory during the execution is concave, while it is convex after the execution of the metaorder.

It is interesting to note that it is possible to derive conditions on the model parameters such that the price follows such a behavior. In particular, it can be proved that for each trade $k = 0,1, \dots, T$, the price trajectory of a buy metaorder during execution ($\delta_v>0$) is concave when

\begin{equation}\begin{split}\label{conccond}
   e_1'\Gamma^{k+1}e_2 > 0,
\end{split}\end{equation}
and convex after the execution for $k = T+1,T+2, \dots, T^{post}$ when 
\begin{equation}\begin{split}\label{convcond}
    e_1'  \Gamma^{k+1} (\tilde{I}_{2p}-\Gamma^T) e_2 < 0.
\end{split}\end{equation}

These conditions can be easily tested on the empirically estimated companion matrix $\Gamma$ of each model.

\section{Solving Volterra Equation for volumes dynamics}\label{VolterraResolution}
Let us solve the Volterra equation for volumes written in Eq.~\eqref{eq: VolterraEqs} by using the solution written in Eq.~\eqref{eq: VolterraSolution}. We obtain
\begin{equation}\begin{split}\label{eq: ExpLaplace}
    &\mathcal{L}\big( f(t) \big) =  \alpha V \mathcal{L}\big[ \theta(T-t) \big] = \alpha V \frac{1-e^{-sT}}{s},\\&
     \mathcal{L}[\mathcal{D}(t)]=\mathcal{L}[e^{-\beta t}] = \frac{1}{s+\beta}.
\end{split}\end{equation}
Note that for $T \rightarrow \infty$, i.e. for infinite metaorders, the Laplace transform of $f(t)$ tends to the case in which the indicator function is always equal to one. 
Thus, we can write the solution in this case as
\begin{equation}\begin{split}
         v(t) = \alpha V  \mathcal{L}^{-1} \bigg[ \frac{\frac{1-e^{-sT}}{s}}{1-\lambda(1/(s+\beta))} \bigg] = \alpha V  \bigg\{ \mathcal{L}^{-1} \bigg[ \frac{s+\beta}{s^2+s(\beta-\lambda)} \bigg] - \mathcal{L}^{-1} \bigg[ \frac{e^{-sT}(s+\beta)}{s^2+s(\beta - \lambda)} \bigg]\bigg\},
\end{split}
\end{equation}

Let us firstly focus on the first term. By calling $k = \beta - \lambda$, we can decompose the fraction into two terms by imposing the relation
\begin{equation}\begin{split}
    \frac{(s+\beta)}{s^2+\beta s - \lambda s} = \frac{A}{s} + \frac{B}{s+k},
\end{split}\end{equation}
which gives $A=\beta/k$ and $B=1-\beta/k$. Thus we would have that
\begin{equation}\begin{split}
 \mathcal{L}^{-1} \bigg[ \frac{s+\beta}{s^2+s(\beta-\lambda)} \bigg] & = \bigg( \frac{\beta}{k} \mathcal{L}^{-1} \bigg[ \frac{1}{s} \bigg] + (1-\frac{\beta}{k}) \mathcal{L}^{-1} \bigg[ \frac{1}{s+k}\bigg] \bigg)\\& = \beta/k  + (1-\beta/k) e^{-k t}
\end{split}\end{equation}

Let us now focus on the second Laplace transform. We have that 
\begin{equation}\begin{split}
    \frac{e^{-sT}(s+\beta)}{s^2+(\beta-\lambda)s} = e^{-sT}\bigg( \frac{1}{s+(\beta-\lambda)} + \frac{\beta}{s(s+(\beta-\lambda)) }\bigg).
\end{split}\end{equation}
Now, we can rewrite the second term as
\begin{equation}\begin{split}\label{eq: Decomposition}
    \frac{1}{s(s+(\beta-\lambda))} = \frac{1/(\beta-\lambda)}{s} - \frac{1/(\beta-\lambda)}{s+(\beta-\lambda)}.
\end{split}\end{equation}
Thus, we obtain that
\begin{equation}\begin{split}
    \mathcal{L}^{-1} & \bigg[ \frac{e^{-sT}(s+\beta)}{s^2+(\beta-\lambda)s}\bigg]  = \mathcal{L}^{-1}\bigg\{ e^{-sT} \bigg[ \frac{1}{s+(\beta-\lambda)} + \beta \bigg( \frac{1/(\beta-\lambda)}{s} - \frac{1/(\beta-\lambda)}{s+(\beta-\lambda)} \bigg) \bigg] \bigg\} \\& = \mathcal{L}^{-1} \bigg[\frac{e^{-sT}}{s+(\beta-\lambda)} \bigg] + \frac{\beta}{\beta-\lambda} \bigg( \mathcal{L}^{-1} \bigg[ \frac{e^{-sT}}{s} \bigg] - \mathcal{L}^{-1} \bigg[ \frac{e^{-sT}}{s+(\beta-\lambda)} \bigg]\bigg),
\end{split}
\end{equation}
where
\begin{equation}\begin{split}
    & \mathcal{L}^{-1} \bigg[\frac{e^{-sT}}{s+(\beta-\lambda)} \bigg] = \theta(t-T)e^{-(\beta-\lambda)(t-T)}\\&
    \mathcal{L}^{-1} \bigg[ \frac{e^{-sT}}{s} \bigg] = \theta(t-T).
\end{split}\end{equation}
The final result for the inverse Laplace transform is given by
\begin{equation}\begin{split}
    \mathcal{L}^{-1} \bigg[ \frac{e^{-sT}(s+\beta)}{s^2+(\beta-\lambda)s}\bigg] = \theta(t-T) \bigg[ \frac{\beta}{\beta-\lambda} + \bigg( 1-\frac{\beta}{\beta - \lambda} \bigg) e^{-(\beta-\lambda)(t-T)} \bigg].
\end{split}\end{equation}
Thus, the final result for the volume before and after the metaorder will be given by
\begin{equation}\begin{split}
    v(t) = \alpha V \bigg\{ \bigg( \frac{\beta}{ \beta - \lambda} + \big(1-\frac{\beta}{ \beta - \lambda}\big) e^{-(\beta - \lambda) t}\bigg) -  \theta(t-T) \bigg( \frac{\beta}{\beta-\lambda} + ( 1-\frac{\beta}{\beta - \lambda} ) e^{-(\beta-\lambda)(t-T)} \bigg)\bigg\}.
\end{split}\end{equation}
Note that this result is always valid for $\beta \neq \lambda$ which would generate a singularity. However, this singularity has not any physical reason to exist. It is just a consequence of the decomposition in Eq.~\eqref{eq: Decomposition} that is only valid for $\beta \neq \lambda$. For $\beta = \lambda$ we would just have 
\begin{equation}\begin{split}
  v(t) & =  \alpha V  \bigg\{ \mathcal{L}^{-1} \bigg[ \frac{s+\beta}{s^2} \bigg] - \mathcal{L}^{-1} \bigg[ \frac{e^{-sT}(s+\beta)}{s^2} \bigg]\bigg\} \\& = \alpha V  \bigg\{ \mathcal{L}^{-1} \bigg[ \frac{1}{s} \bigg] + \beta \mathcal{L}^{-1} \bigg[ \frac{1}{s^2} \bigg] - \mathcal{L}^{-1} \bigg[ \frac{e^{-sT}}{s} \bigg] -\beta \mathcal{L}^{-1} \bigg[ \frac{e^{-sT}}{s^2} \bigg] \bigg\} \\ & = \alpha V \bigg\{ 1 + \beta t - \theta(t-T) - \beta \theta(t-T) (t-T) \bigg\} \\& = \alpha V \bigg\{  1 + \beta t  - \theta(t-T) [ 1 + \beta (t-T) ] \bigg\}.
\end{split}\end{equation}

Thus, the limit case in which $\beta = \lambda$ we would exactly find a linear behaviour.

\section{Price dynamics}\label{PricesDyn}
Using the solution of the volumes dynamics, we can substitute the result in the expression of the prices and obtain the prices dynamics. Let us begin with the $\beta \neq \lambda$ case, for which we obtain
\begin{equation}\begin{split}
p(t) & = \int_0^t G(t-\tau) [v(\tau) + (1-\alpha) V \theta(T-\tau)] d\tau\\& = \int_0^t G(t-\tau) v(\tau) + (1-\alpha) V \int_0^t \theta (T-\tau)  G(t-\tau) d\tau \\ & = \alpha V  \bigg\{ \int_0^t G(t-\tau) \bigg[ \bigg( \frac{\beta}{ \beta - \lambda} + \big(1-\frac{\beta}{ \beta - \lambda}\big) e^{-(\beta - \lambda) \tau}\bigg) \bigg] d\tau \\& - \int_0^t G(t-\tau) \bigg[ \theta(\tau-T) \bigg( \frac{\beta}{\beta-\lambda} + ( 1-\frac{\beta}{\beta - \lambda} ) e^{-(\beta-\lambda)(\tau-T)} \bigg]d\tau \bigg\} \\& + (1-\alpha) V  \int_0^t \theta(T-\tau) G(t-\tau) d\tau.
\end{split}\end{equation}
By taking also in this case an exponential kernel, i.e. $G(t) = e^{-\rho t}$, it is easy to perform the integrals to compute the price dynamics. Indeed, we have that
\begin{equation}\begin{split}\label{eq: IntegralsPrices}
    &\int_0^t e^{-\rho(t-\tau)} e^{-(\beta-\lambda) \tau} d\tau = e^{-\rho t} \int_0^t e^{\tau(\rho - \beta+\lambda)} d\tau = \frac{e^{(-\beta+\lambda) t}-e^{-\rho t}}{\rho - \beta + \lambda};\\& \int_0^t e^{-\rho(t-\tau)} d\tau = \frac{1- e^{-\rho t}}{\rho},\\&
    \int_0^t e^{-\rho(t-\tau)} \theta(T-\tau) d\tau = \frac{e^{-\rho(t-T)}-e^{-\rho t}}{\rho} + \theta(T-t) \frac{1-e^{-\rho(t-T)}}{\rho} = \frac{1-e^{-\rho t}}{\rho} - \theta(t-T)\frac{1 - e^{-\rho (t-T)}}{\rho}, \\& 
    \int_0^t e^{-\rho(t-\tau)} \theta(\tau-T) e^{-(\beta-\lambda)(\tau-t)} d\tau = e^{-\rho t }\theta(t-T) \frac{  e^{-(\beta-\lambda)(t-T) + \rho t } - e^{\rho T} }{\rho - (\beta-\lambda) }.
\end{split}\end{equation}
Thus, the price dynamics for $\beta \neq \lambda$ will be given by
\begin{equation}\begin{split}
    p(t) & = \alpha V \bigg\{  \frac{\beta}{\beta - \lambda} \frac{1-e^{-\rho t}}{\rho} + \bigg( 1-\frac{\beta}{\beta-\lambda} \bigg) \frac{e^{(-\beta + \lambda)t  } - e^{-\rho t}}{\rho - \beta + \lambda} \\& - \theta(t-T)   \bigg( \frac{\beta}{\beta-\lambda} \frac{1 - e^{- \rho(t-T)}}{\rho}  +  ( 1-\frac{\beta}{\beta - \lambda} ) \frac{  e^{-(\beta-\lambda)(t-T)  } - e^{-\rho(t-T)} }{\rho - (\beta-\lambda) } \bigg) \bigg\} \\& + (1-\alpha) V  \bigg\{ \frac{1-e^{-\rho t}}{\rho} - \theta(t-T) \frac{1-e^{-\rho(t-T)}}{\rho}  \bigg\}. 
\end{split}\end{equation}

Let us now consider $\beta = \lambda$. In this case, by substituting the volumes dynamics, we obtain
\begin{equation}\begin{split}
p(t) & = \int_0^t G(t-\tau) \bigg\{ \alpha V \bigg[ 1 + \beta \tau - \theta(\tau-T)\big(1 + \beta (\tau - T)\big) \bigg] + (1-\alpha) V (1-\theta(\tau-T))\bigg\} \\& = \alpha V \bigg\{ \int_0^t e^{-\rho(t-\tau)} d\tau + \beta \int_0^t e^{-\rho(t-\tau)} \tau d\tau - \int_0^t \theta(\tau - T) d\tau - \beta \int_0^t \theta (\tau - T) (\tau-T) e^{-\rho (t-\tau)} d\tau \bigg\} \\& \quad +  (1-\alpha) V  \int_0^t (1 - \theta(\tau-T)) G(t-\tau) d\tau.
\end{split}\end{equation}
Now, we have that
\begin{equation}\begin{split}
& \int_0^t e^{-\rho(t-\tau)} \tau d\tau = \frac{\rho t + e^{-\rho t} - 1}{\rho^2}, \\&
\int_0^t \theta (\tau-T) (\tau - T) e^{-\rho(t-\tau)} d\tau = \theta (t-T) \bigg[ \frac{t-T}{\rho} - \frac{1-e^{-\rho(t-T)}}{\rho^2} \bigg].
\end{split}\end{equation}
Thus, using also the results in Eq.~\eqref{eq: IntegralsPrices}, we find that 
\begin{equation}\begin{split}
p(t) & = \alpha V \bigg\{ \frac{1-e^{-\rho t}}{\rho} + \beta ~ \frac{\rho t + e^{-\rho t}-1}{\rho^2} - \theta(t-T) \bigg[ \frac{1-e^{-\rho(t-T)}}{\rho}+\beta ~ \frac{\rho(t-T)+e^{-\rho(t-T)} - 1}{\rho^2} \bigg] \bigg\} \\& \quad + (1-\alpha) V  \bigg\{ \frac{1-e^{-\rho t}}{\rho} - \theta(t-T) \frac{1-e^{-\rho(t-T)}}{\rho}  \bigg\}.
\end{split}\end{equation}

\section{Modified TIM model discretisation}\label{Appendix_Discretisation}

Let us consider Eq.s~\eqref{eq: VolterraEqs} with $V=0$. To find the discretized version of this model, we divide the continuous time in intervals of length $\Delta t$. Thus, we define a time window of length $t$ as $n \Delta t$, i.e. $t=n \Delta t$, where $n$ represents the number of intervals $\Delta t$ contained in $t$. Now, the discretization process depends on how we construct the rectangular areas within each interval. Specifically, we approximate the area by setting the height of each rectangle to the value of the function evaluated at a chosen point within the interval. This point can be at the beginning, the end, or the midpoint of each interval. Choosing different points for the height results in different forms of the Riemann sum, which subsequently influences the behavior of the discretized model. It can be shown that defining the height of the interval as the beginning point of the interval gives the discretized model corresponding to the Hasbrouck convention, while taking it as the end point gives the TIM model. Thus, both these discrete models tend to the same continuum model, i.e. the one defined in Eq.~\eqref{eq: VolterraEqs}.

Let us start from the volumes and let us discretize the continuous time model in Eq.~\eqref{eq: VolterraEqs}. By approximating the integral with a Riemann sum having using the left endpoints of each subinterval, we can write the volume equation as
\begin{equation}\begin{split}
    v[n\Delta t] = \alpha V \theta(T-n\Delta t)  + \lambda \sum_{k=0}^{t/\Delta t-1} D\big( t - k \Delta t\big) v[k\Delta t] \Delta t. 
\end{split}\end{equation}
If we assume $\Delta t = 1$, the discretization can simply be written as
\begin{equation}\begin{split}\label{eq: DiscrVolumes}
    v_t &= \alpha V \theta(T-n)  + \lambda \sum_{k=0}^{t-1} D\big( t-k \big) v_k =  \alpha V \theta(T-n)  + \lambda \sum_{i=1}^{t}  D\big( i \big) v_{n-i},
\end{split}\end{equation}
where in the second passage we change index in the sum by taking $i=n-k$. It is evident, here, that for $V=0$ we obtain the TIM model volumes. It can be easily shown that taking as discretization point the end of the interval, the Hasbrouck model can be obtained.

The same can be done with the  price equation, 
\begin{equation}\begin{split}
    p[n \Delta t] =  \sum_{k=0}^{t/\Delta t -1 } G[t-i \Delta t] \left(v[k \Delta t] + (1 - \alpha)V \theta(T - k \Delta t)\right) \Delta t,
\end{split}\end{equation}
which, taking $\Delta t = 1$, can be written as
\begin{equation}\begin{split}\label{eq: DiscrPrices}
    p_t =  \sum_{k=0}^{t-1} G[t-k] \left(v_k + (1 - \alpha)V \theta(T - k)\right) = \sum_{i=1}^{n} G[i] \left(v_{t-i} + (1 - \alpha)V \theta\big(T - (n-i)\big)\right),
\end{split}\end{equation}
where the coefficients $G[i]$  correspond to the TIM coefficients. Also in this case, taking as discretization point the end of the interval, one obtains the Hasbrouck price model.

\begin{figure}[!htb]
\centering
\includegraphics[width=8cm]{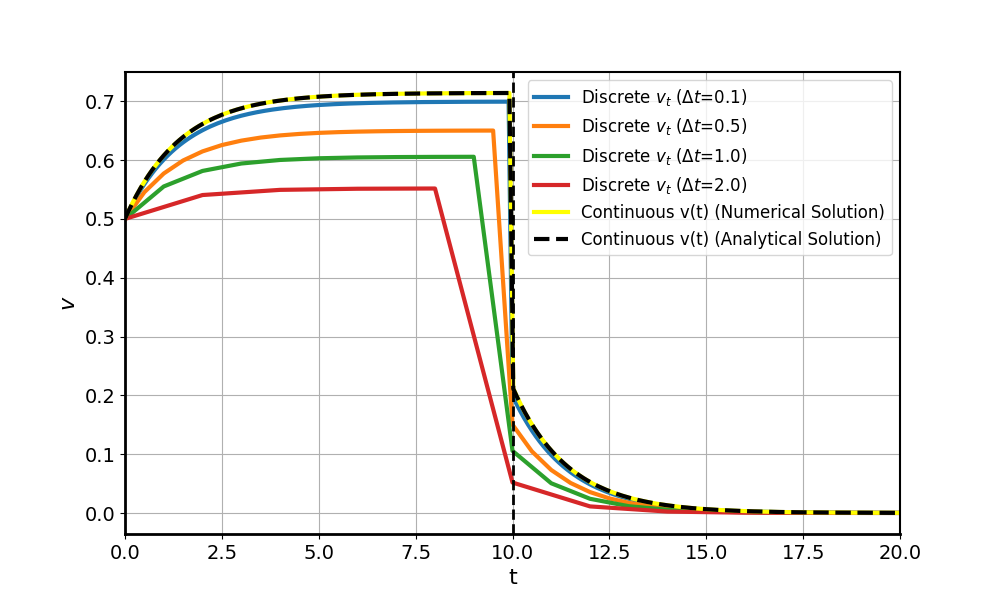}
\includegraphics[width=8cm]{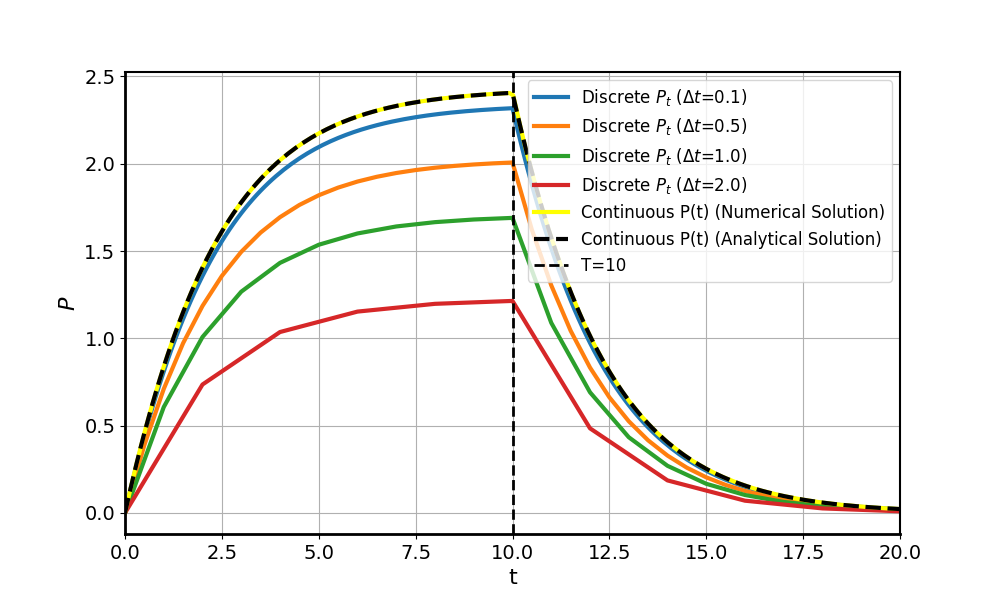}
\caption{Left panel. Comparison between the volume dynamics in the discrete and continuous time model in the exponential kernel case for $\alpha$ = 1.0, $V$ = 1.0, $\lambda$ = 0.4, $\beta$ = 0.8, $T$ = 10.
Right panel. Comparison between the price dynamics in the discrete and continuous time model in the exponential kernel case for $\alpha$ = 0.5, $V$ = 1.0, $\lambda$ = 0.3, $\rho=0.5$, $\beta$ = 1, $T$ = 5. }
\label{fig:Cont_Lim_V}
\end{figure}

We test numerically the convergence of the discrete time model to the one in continuous time.
Fig.~\ref{fig:Cont_Lim_V}  
shows the numerical solution of the two models for volume and price dynamics, in the exponential kernel case for which we also have the analytical solution. Both figures clearly show the convergence of the discrete solutions to the continuum ones in the limit $\Delta t \rightarrow 0$. 

\section{Diffusivity of price}\label{Diffusivity}
Here we consider a discrete time stationary generalization of our model where many metaorders are executed. Let $V_t$ be the volume traded at time $t$ due to child orders of the active metaorders and let $v_t$ represent the volume of other orders. For the latter we assume that, in the absence of metaroders,  it can be described as a short memory process, namely a stationary autoregressive process of finite order $p$. We can rewrite the second equation of (\ref{eq: VolterraEqs}) in discrete time as
\begin{equation}\begin{split}\label{eq:VolumeLMF}
    v_t = \sum_{i=1}^{p} d_i\, v_{t-i} + \big(\alpha V_t + \epsilon_t\big),
\end{split}\end{equation}
where we added a white noise term $\epsilon_t$. 
According to the Lillo–Mike–Farmer (LMF) model and to empirical evidences, $V_t$  is a long memory process, meaning that
\begin{equation}\begin{split}
    r_V(\tau) = \text{Cov}(V_t, V_{t+\tau}) \sim \tau^{-\gamma}, \quad \text{as } \tau\rightarrow \infty,\quad 0<\gamma < 1. 
\end{split}\end{equation}
We are going to show that the autocorrelation of $v_t$ is also long memory with the same exponent $\gamma$.
To this end, we  rewrite the volume equation as
\begin{equation}\begin{split}
v_t= \sum_{i=1}^{p} d_i\, v_{t-i} + \eta_t,
  \label{eq:arp}
\end{split}\end{equation}
where $\eta_t = \epsilon_t + \alpha V_t$ is also a long-memory process with asymptotic autocovariance
\begin{equation}\begin{split}
  r_\eta(\tau) \;=\; \operatorname{Cov}(\eta_t,\eta_{t+\tau})
  \;\sim\; C_\eta\, \tau^{-\gamma}, 
  \qquad \text{as }\tau\to\infty,\ \ 0<\gamma<1.
  \label{eq:longmemory}
\end{split}\end{equation}
Let us now introduce the lag operator $L$ as $L v_t = v_{t-1}$ and the polynomial
\begin{equation}\begin{split}
  \phi(L) \;=\; 1 - \sum_{i=1}^{p} d_i L^i .
\end{split}\end{equation}
Then \eqref{eq:arp} can be written compactly as
\begin{equation}\begin{split}
  \phi(L)\, v_t \;=\; \eta_t.
\end{split}\end{equation}
Now, if the process is stationary, i.e. all the roots of $\phi(z)$ lie outside the unit circle, we can rewrite it as
\begin{equation}\begin{split}
  v_t \;=\; \psi(L)\,\eta_t 
  \;=\; \sum_{k=0}^{\infty} h_k\, \eta_{t-k},
  \qquad \text{where } \psi(L)=\phi(L)^{-1}.
\end{split}\end{equation}
Let us now compute its covariance
\begin{equation}\begin{split}
    r_v(\tau) & = \text{Cov}\big(\sum_k h_k\eta_{t-k}, \sum_\ell h_\ell \eta_{t+\tau-\ell} \big) \\ & = \sum_{k,l }h_k h_\ell~  \text{Cov}(\eta_{t-k}, \eta_{t+\tau-\ell}) \\&= \sum_{k,\ell}h_k h_\ell ~ r_\eta(\tau + k -\ell).
\end{split}\end{equation}
Let us now define the Fourier transform as
\begin{equation}\begin{split}
    f_X(\lambda) = \frac{1}{2\pi} \sum_{\tau \in \mathcal{Z}} r_X(\tau)e^{-i\tau \lambda}.
\end{split}\end{equation}
Thus, we can find that
\begin{equation}\begin{split}
    f_v(\lambda) &= \frac{1}{2\pi} \sum_{\tau}\sum_{k,\ell} h_k h_\ell r_\eta (\tau + k - \ell) e^{-i\tau \lambda}\\& = \sum_{k,\ell} h_k h_\ell  e^{-i(\ell-k) \lambda} \bigg[ \frac{1}{2\pi} \sum_s r_\eta(s) e^{-is\lambda} \bigg] \\& = \bigg( \sum_k h_k e^{i k\lambda}\bigg) \bigg( \sum_\ell h_\ell e^{-i\ell \lambda} \bigg) \bigg[ \frac{1}{2\pi} \sum_s r_\eta(s) e^{-is\lambda} \bigg] \\& = \bigg| H(e^{-i\lambda}) \bigg|^2 f_\eta(\lambda),
\end{split}\end{equation}
where we defined the transfer function as
\begin{equation}\begin{split}
    H(e^{-i\lambda}) = \psi(e^{-i\lambda}) = \frac{1}{\phi(e^{-i\lambda})}.
\end{split}\end{equation}
Thus, we find the spectrum of $v_t$ as
\begin{equation}\begin{split}
  f_v(\lambda) \;=\; \frac{f_\eta(\lambda)}{\lvert \phi(e^{-i\lambda})\rvert^2},
  \qquad \lambda\in[-\pi,\pi].
\end{split}\end{equation}
If $\eta_t$ has long memory with $r_\eta(\tau)\sim C_\eta \tau^{-\gamma}$, then as $\lambda\to 0$,
\begin{equation}\begin{split}
  f_\eta(\lambda) \;\sim\; C'_\eta\, |\lambda|^{\,\gamma-1}
  \quad\Longrightarrow\quad
  f_v(\lambda) \;\sim\; 
  \frac{C'_\eta\, |\lambda|^{\,\gamma-1}}{\lvert \phi(e^{-i\lambda})\rvert^2}.
\end{split}\end{equation}
Furthermore, as $\lambda \rightarrow 0$ we also have
\begin{equation}\begin{split}
    \phi(e^{-i\lambda}) \sim \phi(1)
\end{split}\end{equation}
from which
\begin{equation}\begin{split}
    f_v(\lambda) \sim \frac{C'_\eta\, |\lambda|^{\,\gamma-1}}{\lvert \phi(1)\rvert^2}.
\end{split}\end{equation}
This implies that in our model the volume is still a long memory process. In fact, if we come back in the time space we find that
\begin{equation}\begin{split}
    r_v(\tau) \sim \frac{C_\eta \tau^{-\gamma}}{|\phi(1)|^2}.
\end{split}\end{equation}
Thus, when the metaorder volume \(V_t\) is long memory (as for example in the LMF model), the volume \(v_t\) is long-memory, and the usual diffusivity conditions of the standard propagator model apply unchanged \cite{bouchaud2009markets}. In fact, Equation~\eqref{eq:VolumeLMF} exactly corresponds to the sum of the terms $\tilde{v}$ and $V$ that appear in the propagator relation in Equation~\eqref{eq:PriceDecomposition}.

\end{document}